\documentclass[twocolumn,twocolappendix]{aastex702}
\usepackage{float}
\usepackage{graphicx}
\usepackage{amsmath}
\usepackage{natbib}
\usepackage{color}
\usepackage{verbatim}
\usepackage[utf8]{inputenc}
\usepackage{mathtools}
\usepackage{amssymb}
\usepackage{textcomp}
\usepackage{enumitem}
\usepackage{braket}
\usepackage{graphicx} 
\usepackage{tablefootnote,tabularx}
\usepackage{subfigure}
\usepackage{siunitx}

\begin{document}

\title{Discovery, Characterization, and Potential Origins of a Stream in the Stellar Halo of Nearby LMC-Mass Galaxy NGC~55}

\author[0009-0001-1147-6851]{Benjamin N.\ Velguth}
\altaffiliation{NSF Graduate Research Fellow}
\email[show]{benjamin.n.velguth.gr@dartmouth.edu}
\affiliation{Department of Physics and Astronomy, Dartmouth College, Hanover, NH 03755, USA}

\author[0000-0001-9649-4815]{Bur\c{c}in Mutlu-Pakdil}
\email{Burcin.Mutlu-Pakdil@dartmouth.edu}
\affiliation{Department of Physics and Astronomy, Dartmouth College, Hanover, NH 03755, USA}

\author[0000-0002-9599-310X]{Erik Tollerud}
\email{etollerud@stsci.edu}
\affiliation{Space Telescope Science Institute, 3700 San Martin Drive, Baltimore, MD 21218, USA}

\author[0000-0003-0256-5446]{Sarah Pearson}
\affiliation{Technical University of Denmark, Elektrovej 327, 2800 Kgs. Lyngby, Denmark}
\affiliation{DARK, Niels Bohr Institute, University of Copenhagen, Jagtvej 155A, 2200 Copenhagen,  Denmark}
\email{sapea@dtu.dk}

\author[0000-0002-0956-7949]{Kristine Spekkens}
\email{kristine.spekkens@queensu.ca}
\affiliation{Department of Physics, Engineering Physics and Astronomy, Queen’s University, Kingston, ON K7L 3N6, Canada}

\author[0000-0001-9775-9029]{Amandine Doliva-Dolinsky}
\email{amandinedolinsky@gmail.com}
\affil{Department of Physics, University of Surrey, Guildford, Surrey GU2 7XH, UK}

\author[0000-0002-5434-4904]{Michael G.\ Jones}
\email{mgjones@ipac.caltech.edu}
\affiliation{IPAC, Mail Code 100-22, Caltech, 1200 E. California Boulevard, Pasadena, CA 91125, USA}

\author[0000-0002-1763-4128]{Denija Crnojevi\'{c}}
\email{dcrnojevic@ut.edu}
\affil{Department of Physics \& Astronomy, University of Tampa, 401 West Kennedy Boulevard, Tampa, FL 33606, USA}

\author[0000-0002-3936-9628]{Jeffrey L.\ Carlin}
\email{jeffreylcarlin@gmail.com}
\affil{AURA/Rubin Observatory, 950 North Cherry Avenue, Tucson, AZ 85719, USA} 

\author[0000-0003-4102-380X]{David J.\ Sand}
\email{dsand@arizona.edu}
\affiliation{Department of Astronomy/Steward Observatory, 933 North Cherry Avenue, Room N204, Tucson, AZ 85721-0065, USA}

\author[0000-0001-8354-7279]{Paul Bennet}
\affiliation{Space Telescope Science Institute, 3700 San Martin Drive, Baltimore, MD 21218, USA}
\email{pbennet@stsci.edu}

\author[0000-0003-1479-3059]{Guy Stringfellow}
\email{Guy.Stringfellow@colorado.edu}
\affiliation{Center for Astrophysics and Space Astronomy, University of Colorado Boulder, Boulder, CO 80309, USA}

\author[0000-0003-1697-7062]{William Cerny}
\email{william.cerny@yale.edu}
\affiliation{Department of Astronomy, Yale University, New Haven, CT 06520, USA}

\author[0000-0001-8251-933X]{Alex Drlica-Wagner}
\email{kadrlica@fnal.gov}
\affiliation{Department of Astronomy and Astrophysics, University of Chicago, Chicago, IL 60637, USA}
\affiliation{Fermi National Accelerator Laboratory, P.O.\ Box 500, Batavia, IL 60510, USA}
\affiliation{Kavli Institute for Cosmological Physics, University of Chicago, Chicago, IL 60637, USA}
\affiliation{NSF-Simons AI Institute for the Sky (SkAI),172 E. Chestnut St., Chicago, IL 60611, USA}

\author[0000-0002-1594-1466]{Joanna D.\ Sakowska}
\email{jsakowska@iaa.es}
\affiliation{Instituto de Astrof\'isica de Andaluc\'ia (CSIC), Glorieta de la Astronom\'ia,  E-18080 Granada, Spain}

\author[0000-0002-4350-7632]{Jaclyn Jensen}
\email{jaclyn.r.jensen@dartmouth.edu}
\affiliation{Department of Physics and Astronomy, Dartmouth College, Hanover, NH 03755, USA}

\author[0000-0002-4350-7632]{Laura C.\ Hunter}
\email{laura.c.hunter@dartmouth.edu}
\affiliation{Department of Physics and Astronomy, Dartmouth College, Hanover, NH 03755, USA}

\author[0000-0002-5564-9873]{Eric F.\ Bell}
\email{ericbell@umich.edu}
\affiliation{Department of Astronomy, University of 
Michigan, 323 West Hall, 1085 S. University Ave., Ann Arbor, MI, 48105-1107, USA} 

\author[]{David Mart\'{i}nez-Delgado}
\email{dmartinez@iaa.es}
\affiliation{Centro de Estudios de F\'isica del Cosmos de Arag\'on (CEFCA),
Unidad Asociada al CSIC, Plaza San Juan 1, 44001 Teruel, Spain}
\affiliation{ARAID Foundation, Avda. de Ranillas, 1-D, E-50018 Zaragoza,
Spain}

\author[0000-0002-7155-679X]{Anirudh Chiti}
\email{achiti@stanford.edu}
\affiliation{Kavli Institute for Particle Astrophysics \& Cosmology, P.O. Box 2450, Stanford University, Stanford, CA 94305, USA} 

\author[0000-0003-2599-7524]{Adam Smercina}
\email{Adam.Smercina@tufts.edu}
\affiliation{Department of Physics and Astronomy, Tufts University, 574 Boston Ave., Medford, MA 02155, USA}

\author[0000-0002-3204-1742]{Nitya Kallivayalil}
\email{njk3r@virginia.edu}
\affiliation{Department of Astronomy, University of Virginia, 530 McCormick Road, Charlottesville, VA 22904, USA}

\author[0000-0002-8217-5626]{Deepthi S. Prabhu}
\affiliation{Steward Observatory, University of Arizona, 933 North Cherry Avenue, Tucson, AZ 85721-0065, USA}
\email{dprabhu@arizona.edu}

% check back on email 
\author[0000-0001-5805-5766]{Alexander H. Riley}
\email{alexander.riley@fysik.lu.se}
\affiliation{Institute for Computational Cosmology, Department of Physics, Durham University, South Road, Durham DH1 3LE, UK}
\affiliation{Lund Observatory, Division of Astrophysics, Department of Physics, Lund University, SE-221 00 Lund, Sweden}

\author[0000-0003-1680-1884]{Yumi Choi}
\email{yumi.choi@noirlab.edu}
\affiliation{NSF NOIRLab, 950 N. Cherry Ave., Tucson, AZ 85719, USA}

\author[0000-0002-8282-469X]{Noelia E. D. No\"{e}l}
\email{n.noel@surrey.ac.uk}
\affiliation{Department of Physics, University of Surrey, Guildford GU2 7XH, UK}

\author[0009-0007-9488-7050]{Sasha N. Campana}
\email{sasha.n.campana.gr@dartmouth.edu}
\affiliation{Department of Physics and Astronomy, Dartmouth College, Hanover, NH 03755, USA}

\author[0000-0003-4394-7491]{Guinevere Herron}
\email{guinevere.herron.gr@dartmouth.edu}
\affiliation{Department of Physics and Astronomy, Dartmouth College, Hanover, NH 03755, USA}

\author[0000-0002-7123-8943]{Alistair R. Walker}
\email{alistair.walker@noirlab.edu}
\affiliation{Cerro Tololo Inter-American Observatory/NSF NOIRLab, Casilla 603, La Serena, Chile}

\author[0000-0002-6021-8760]{Andrew B. Pace}
\email{pvpace1@gmail.com}
\affiliation{Department of Astronomy, University of Virginia, 530 McCormick Road, Charlottesville, VA 22904, USA}

\author[0000-0002-9144-7726]{Clara E. Mart\'inez-V\'azquez}
\email{clara.martinez@noirlab.edu}
\affiliation{NSF NOIRLab, 670 N. A'ohoku Place, Hilo, Hawai'i, 96720, USA}

\author[0000-0003-4383-2969]{Clecio R. Bom}
\email{debom@cbpf.br}
\affiliation{Centro Brasileiro de Pesquisas F\'isicas, Rua Dr. Xavier Sigaud 150, 22290-180 Rio de Janeiro, RJ, Brazil}

\author[0000-0002-3690-105X]{Julio A. Carballo-Bello}
\email{jcarballo@academicos.uta.cl}
\affiliation{Instituto de Alta Investigaci\'on, Universidad de Tarapac\'a, Casilla 7D, Arica, Chile}

\author[0000-0002-8093-7471]{Pol Massana}
\email{pol.massana@noirlab.edu}
\affiliation{NSF NOIRLab, Casilla 603, La Serena, Chile}

\author[0000-0003-0105-9576]{Gustavo E. Medina}
\email{gustavo.medina@utoronto.ca}
\affiliation{David A. Dunlap Department of Astronomy \& Astrophysics, University of Toronto, 50 St George Street, Toronto ON M5S 3H4, Canada}
\affiliation{Department of Astronomy and Astrophysics, University of Toronto, 50 St. George Street, Toronto ON, M5S 3H4, Canada}

% \author[]{Others}
% \email{}
% \affiliation{}

\begin{abstract}
We present a previously undetected stellar stream in the halo of the LMC-mass dwarf galaxy NGC~55 (2 Mpc), as part of an ongoing effort to characterize the stellar halos of LMC/SMC-mass dwarfs in the DEEP component of the DECam Local Volume Exploration (DELVE) survey. This structure is aligned with an outflow traced by H$\alpha$ emission, a spur and cloud of neutral hydrogen, and the ultra-diffuse and possibly disrupting satellite galaxy NGC~55-dw1. We investigate possible \emph{ex-situ} progenitor scenarios for this stream through the local luminosity-metallicity relation and a set of toy dynamical models that include gaseous, stellar, and dark matter components. If the stream is a product of a previous merger, the progenitor galaxy had an absolute magnitude $M_V \leq -7.2$ and a stellar mass $M_* \geq 10^5 M_{\odot}$ based on extrapolations of the detected stellar populations. If it is disrupted material from the ultra-diffuse satellite, their shared progenitor had an absolute magnitude $M_V \leq -8.3$ and a stellar mass $M_* \geq 10^{5.2} M_{\odot}$. We compare our stream and progenitor scenarios to those for NGC~300, a galaxy with a stellar mass similar to that of NGC~55 and known to host multiple structures in its halo. Our results demonstrate that even relatively isolated LMC-mass galaxies can have complex accretion histories, providing new tests of hierarchical galaxy formation at low masses.
    
\end{abstract}


% \uat{Galaxy Mergers}{608} ---

\section{Introduction}
\label{sec:intro}

Most galaxies contain extended, low surface-brightness stellar halos. Around more massive systems, they originate from accreted satellites, serving as direct evidence of the hierarchical assembly of galaxies predicted by $\Lambda$ Cold Dark Matter ($\Lambda$CDM) cosmology \citep[e.g.][]{toomre_galactic_1972,white_core_1978}. Far out in the gravitational potential where dynamical timescales are long, these accreted stellar populations retain the properties of their progenitors \citep[e.g.][]{fardal_investigating_2007,law_sagittarius_2010,pop_galaxies_2017} and are sensitive tracers of the shape and dark matter distribution of the host halo \citep[e.g.][]{fardal_inferring_2013,foster_kinematics_2014,pearson_mapping_2022,nibauer_constraining_2023,walder_probing_2025,nibauer_testing_2026,starkman_stream_2025,chemaly_hierarchical_2026}. As a result, stellar halos are one of the primary ways we can understand galaxy formation and cosmology in the nearby universe \citep[e.g.][]{bullock_tracing_2005}. 

Halos primarily built from accreted satellites, or \emph{ex-situ} stellar halos, have been observed around galaxies like the Milky Way \citep[MW, e.g.][]{bell_accretion_2008,shipp_stellar_2018,helmi_streams_2020}, M31 \citep[e.g.][]{ibata_giant_2001,dsouza_andromeda_2018,escala_elemental_2021}, M81 \citep{smercina_saga_2020}, M82 \citep{velguth_timeline_2024}, M94 \citep{gozman_saying_2023}, and Cen A \citep{crnojevic_extended_2016,aghdam_complex_2024}, but it is unclear whether or not dwarf galaxies ($\text{log} \ M_*/M_\odot \lesssim 10.0$) assemble their stellar halos in the same manner as their more massive counterparts. While all galaxies are predicted to accrete smaller galaxies, it is possible that the stellar halos of dwarf galaxies are predominantly \emph{in-situ}, that is, composed of stars from the inner regions that get kicked up through dynamical interactions or supernovae-driven radial migration \citep{stinson_feedback_2009,el-badry_breathing_2016}. 
% as there may not be enough faint galaxies to build up a dwarf stellar halo through mergers. 
% It is possible that the stellar halos of dwarf galaxies are predominantly \emph{in-situ}, that is, composed of stars from the inner regions that get kicked up through dynamical interactions or supernovae-driven radial migration \citep{stinson_feedback_2009,el-badry_breathing_2016}. 

% It is likely that stellar halos in this host mass range are a combination of \emph{in-} and \emph{ex-situ} (CITATIONS), and it is possible that there are environmental metallicity dependencies for what formation pathway plays the larger role (CITATIONS). 

Both ex-situ and in-situ mechanisms have been proposed to contribute to the formation of dwarf-galaxy stellar halos, with observational and theoretical evidence supporting each channel. Several nearby dwarfs show direct signatures of accretion: NGC~4449 is accreting one of its dwarf satellites \citep{martinez-delgado_dwarfs_2012}, and DDO~44 is actively being disrupted by its host, NGC~2403 \citep{carlin_tidal_2019}. NGC~300 has recently been found to host multiple coherent structures in its stellar halo \citep{fielder_streams_2025}, and resolved-star observations with JWST of the LMC-mass galaxy Ark~277 revealed ``shelves" in its outskirts consistent with accretion origins \citep{conroy_detection_2023}. Streams, asymmetric halos, and an abundance of shell-like structures have been found around many nearby dwarf galaxies in integrated-light observations \citep{sakowska_stellar_2026}. Dark matter-only simulations combined with empirical galaxy models similarly predict \emph{ex-situ} contributions \citep[][]{deason_dwarf_2022,cooper_simulations_2025}, with extended stellar halos forming predominantly through intermediate-mass (mass ratio of $\sim$1:5) mergers. In the ultra-faint dwarf regime (luminosity $\lesssim 10^5 L_\odot$), simulations predict that their asymmetric outskirts are due to the late-time accretion of even smaller companions \citep[][]{goater_edge_2023}, showing that \emph{ex-situ} halos form around some of the smallest galaxies in $\Lambda$CDM. On the other hand, integrated-light observations of dwarfs at a variety of masses within the Local Volume show smooth, spheroidal shapes and mixed populations in their outskirts, indicative of \emph{in-situ} halo formation channels \citep{kado-fong_tracing_2020,hunter_ultra-deep_2026}. Cosmological hydrodynamic simulations also find substantial \emph{in-situ} contributions: FIRE predicts predominantly \emph{in-situ} stellar halos in dwarfs \citep{kado-fong_situ_2022}, while Auriga predicts that \emph{in-situ} stars dominate the inner halos of higher-mass dwarfs ($\text{log} \ M_*/M_\odot \gtrsim 9.0$) and can dominate at all radii in lower-mass systems ($\text{log} \ M_*/M_\odot \lesssim 8.0$, \citealt{tau_role_2025}). Many observational studies find extended stellar components of dwarf galaxies at various masses with ambiguous origins \citep[e.g.][]{annibali_smallest_2020,jensen_small-scale_2023,sacchi_smallest_2024,annibali_euclid_2026}. The relative importance of \emph{in-} and \emph{ex-situ} formation channels, and how it varies with dwarf-galaxy mass, metallicity, and assembly history, remains an open question in near-field cosmology.

% Outside of the handful of confirmed accretion-built halos, many observational studies find extended stellar components of dwarf galaxies at various masses with ambiguous origins \citep[e.g.][]{annibali_smallest_2020,jensen_small-scale_2023,sacchi_smallest_2024,annibali_euclid_2026}.

%Analysis of the FIRE cosmological zoom-in simulations predicts that the stellar halos of dwarfs form predominantly \emph{in-situ} \citep{kado-fong_situ_2022}, and analysis of the Auriga simulations \citep{tau_role_2025} predict \emph{in-situ} stars within 6 half-light radii for high-mass dwarfs ($\text{log} \ M_*/M_\odot \gtrsim 9.0$) and at all radii for lower-mass dwarfs ($\text{log} \ M_*/M_\odot \lesssim 8.0$). Clearly, understanding how dwarf galaxies acquire their stellar halos is an unsolved question in near-field cosmology. 

% Recent observations of Milky Way satellites uncovered a low-density outer stellar envelope with member stars sometimes out to 10 half-light radii which could be evidence of accreted populations \citep{jensen_small-scale_2023} % sorry jax

One of the most direct ways to address this problem is through studies of the resolved stars in the halos of nearby dwarf galaxies, which can reveal low-surface-brightness structures and constrain the origins of their stellar populations. Only a handful of dwarf galaxies are sufficiently nearby for individual halo stars to be resolved from the ground. The DEEP component of the DECam Local Volume Exploration (DELVE) survey \citep{drlica-wagner_decam_2021,drlica-wagner_decam_2022} was designed to map four nearby, isolated Magellanic Cloud-mass galaxies out to their virial radii, enabling searches for both satellites \citep{medoff_delve-deep_2025,doliva-dolinsky_ngc_2025} and stellar halo substructures \citep{fielder_streams_2025} through resolved stellar populations. This survey is complemented by the Magellanic Analog Dwarf Companions and Stellar Halos (MADCASH) survey \citep{carlin_first_2016,carlin_census_2024}, which targets an additional seven nearby dwarf galaxies with Subaru/Hyper Suprime-Cam \citep{miyazaki_hyper_2012} and DECam \citep{flaugher_dark_2015}. 

NGC~55 is one of the systems observed by DELVE-DEEP. It is a barred spiral galaxy at $\sim$2~Mpc \citep{de_vaucouleurs_third_1991,gieren_araucaria_2008} with a stellar mass \citep[$M_* = 3\times 10^9 M_\odot$,][]{mcconnachie_observed_2012,dooley_observers_2017} comparable to that of the LMC. It is a weakly bound member at the outskirts of the Sculptor Group \citep{karachentsev_local_2005}. Its relative isolation \citep[aside from its two confirmed satellites,][]{medoff_delve-deep_2025} and high inclination make it an ideal target for resolved-star searches for stellar halo substructure. This galaxy has long been hypothesized to have experienced a merger or interaction \citep{hummel_neutral_1986,puche_h_1991} based on asymmetries in its stellar and gas disk. 
Observations from both the ground \citep[Subaru/Suprime Cam,][]{tanaka_structure_2011} and space \citep[Hubble Space Telescope,][]{mouhcine_halos_2005} have also revealed old, metal-poor stellar populations in its outskirts. Together with the disturbed morphology and kinematics of its gaseous disk \citep{westmeier_gas_2013}, these observations motivate an \emph{ex-situ} contribution to the formation of NGC~55's stellar halo. The recent discovery of the extremely diffuse galaxy NGC~55-dw1 \citep[hereafter dw1,][]{mcnanna_search_2024} confirmed as a satellite of NGC~55 through distance association provides further evidence for a complex interaction history. This satellite is old, metal-poor, and extremely extended; to date, dw1 is the most diffuse galaxy known at its luminosity $(M_V \approx -8)$, suggesting that it may have been disturbed by NGC~55 \citep{mcnanna_search_2024,medoff_delve-deep_2025}.

Here we present a previously undetected stream-like structure in the northern halo of NGC~55 that extends in the direction of dw1 and is aligned with multiple displaced neutral hydrogen features and an outflow traced by H$\alpha$ emission. We investigate possible formation pathways for these structures and whether they may share a common origin with dw1. %their potentially shared origins with dw1. 
Section \ref{sec:dandm} outlines the data analyzed in this work, with Section \ref{subsec:delvedeep} describing the DELVE-DEEP observations used to study the stellar populations of NGC~55, and Section \ref{subsec:hi} showing the {\sc Hi} data originally presented in \cite{westmeier_gas_2013} that we use in tandem with the resolved stars to investigate merger history. Section \ref{sec:mf} describes our implementation of a matched filter detection algorithm, and Section \ref{sec:results} shows the detection of the stream in the stellar density maps as well as its association with dw1, the outflow, and disturbed {\sc Hi} gas. Section \ref{sec:discussion} provides an overview of previous studies of NGC~55's stellar halo (Section \ref{subsec:prev-stud}), explores three potential \emph{ex-situ} formation channels for the stellar stream (Sections \ref{subsec:ind-prog}, \ref{subsec:dynamics}), compares our results to those found by \cite{fielder_streams_2025} in the halo of NGC~300 (Section \ref{subsec:compn300}), and finally describes the limitations of our study and avenues for future work (Section \ref{subsec:limits-future}). We conclude in Section \ref{sec:conclusion}.

\section{Observations}% and Data Reduction}
\label{sec:dandm}

% \begin{deluxetable*}{cccccc}[bth]
% \tablecaption{
%     \textnormal{Properties of NGC 55 and its satellite NGC 55-dw1 }
%     \label{Tab:table}
% }
% \tablecolumns{6}
% \setlength{\extrarowheight}{4pt}
% \tablewidth{\linewidth}
% \tabletypesize{\small}
% \tablehead{
% \colhead{Galaxy} & 
% \colhead{Ra. (Deg.)} &  
% \colhead{Dec. (Deg.)} &
% \colhead{$m - M$ (mag)} &
% \colhead{Distance (Mpc)} &
% \colhead{Stellar Mass $(M_\odot)$} &
% }
% \startdata
% \hline
% NGC~55 & $00^{\text{h}}14^{\text{m}}53.6^{\text{s}}$ & \ang{-39; 11; 47.9} & 26.61 & 2.1 & $3 \times 10^9$  \\ 
% NGC~55-dw1 & $00^{\text{h}}15^{\text{m}}28.8^{\text{s}}$  & \ang{-38; 25; 8.4} & 26.71 & 2.2 & $1.42 \times 10^5$  \\
% \hline
% \enddata
% \tablecomments{ References with footnotes here for each property  
% }

% \end{deluxetable*}

We use the DEEP component of the DELVE survey to identify stellar halo substructure around NGC~55 and compare with {\sc Hi} observations to search for signs of recent dynamical interactions. Here we describe the photometry used from the DELVE-DEEP survey (Section~\ref{subsec:delvedeep}) and the {\sc Hi} observations used here (Section~\ref{subsec:hi}).

\subsection{The DELVE-DEEP Survey}
\label{subsec:delvedeep}
% Paraphrase from Jonah paper

The DELVE survey uses the Dark Energy Camera \citep{flaugher_dark_2015} on the Blanco 4m Telescope at the Cerro Tololo Inter-American Observatory \citep{drlica-wagner_decam_2021,drlica-wagner_decam_2022}. This three-component survey aims to observe dwarf galaxies across multiple environments within the Local Volume. NGC~55 was observed as part of DELVE-DEEP, the survey component designed to target nearby Magellanic Cloud-mass ($M_* \approx 10^8 - 10^9 M_\odot$) dwarf galaxies and characterize their stellar halos and satellite populations.

% This survey runs in parallel with the MADCASH Survey \citep{carlin_first_2016}, which observes an additional seven nearby dwarf galaxies with wide field coverage using Subaru Hyper-Suprime Cam. These surveys offer similar depths to various LSST data releases (DELVE: Year 2 and MADCASH: Year 10) providing an early glimpse into the stellar halos and satellite systems that will be discovered in the coming years. 

NGC 55 was observed on seventeen nights between August 2019 and July 2021. To cover the system out to approximately its virial radius ($\sim120$~kpc),
%To roughly cover the entire virial radius of this system ($\sim$120 kpc), 
14 fields spanning a radius of 3.25 degrees were imaged in the \emph{g} and \emph{i} bands. For each field, 12$\times$300~s \emph{g}-band and 7$\times$300~s \emph{i}-band exposures were taken with good atmospheric conditions ($<$1\arcsec\ seeing). These observations were combined with $\sim 10 \times 90 \text{s}$ exposures from the Dark Energy Survey \citep[DES, ][]{abbott_dark_2018,abbott_dark_2021}. We use the Single Object Fitting (SOF) catalogs generated from this data; see \cite{tan_pride_2025} for a description of data reduction and photometric measurement pipeline.
% Within a degree of NGC~55, the observations achieve an average $10\sigma$ depths (defined as the PSF magnitude at which the signal-to-noise ratio is equal to 10) of $g = 25.9$ and $i = 24.5$ and $5\sigma$ depths of $g = 26.6$ and $i = 25.1$ \citep{mcnanna_search_2024}. 
Additional artificial star tests show that the coadded DELVE-DEEP observations are 90\% complete down to $\sim$25.3 in the \emph{g}-band and $\sim$24.6 in the \emph{i}-band. At the distance of NGC~55, these data provide reliable photometry to roughly 1.5 magnitudes below the tip of the red giant branch \citep[TRGB, $m_g \approx 24.3$ and $m_i \approx 22.8$;][]{medoff_delve-deep_2025}. These completeness limits provide a conservative estimate of the effective depth of the SOF catalogs used in this work.
%which is a good but pessimistic proxy for the depths achieved in the Single Object Fitting (SOF) catalogs used here. 
We correct for Galactic extinction using the \cite {schlegel_maps_1998} dust maps and the DES extinction coefficients \citep{abbott_dark_2021}. %The data reduction and photometric measurement pipeline used for the data we analyze here is described in \cite{tan_pride_2025}.

In ground-based, seeing-limited observations, unresolved background galaxies are the primary impediment to resolved-star observations of low surface brightness features in the halos of nearby galaxies. At faint magnitudes, these sources can dominate the number counts in resolved-star catalogs and, because they are spatially clustered, can mimic genuine stellar overdensities.
%These sources are often clustered and dominate the number counts of resolved-star catalogs. 
To mitigate this contamination, we identify and remove likely background sources using the DES-Y6 GOLD source classifier \citep[following][]{fielder_streams_2025,bechtol_dark_2026}. 
The pipeline uses the size and signal-to-noise ratio of each source to assign an ``extended-class'' value through interpolation in this two-dimensional parameter space. These integer values range from 0 (very likely a star) to 4 (very likely a galaxy). Following \citet{fielder_streams_2025}, we keep sources with extended-class values of 0, 1, or 2 and exclude those with higher values as likely background contaminants.

\subsection{Neutral Hydrogen Observations}
\label{subsec:hi}

To complement our resolved star search for halo substructure, we additionally use {\sc Hi} data first presented in \cite{westmeier_gas_2013}. These data were taken with the Australian Telescope Compact Array (ATCA) between November 2007 and February 2009, consisting of 32 pointings mosaicked into a 2 deg$^2$ ($\sim$65 kpc$^2$ at a distance of $\sim2$ Mpc) field of view centered on NGC~55, with a beam size of 158 $\times$ 84 arcsec. These data have a velocity coverage of roughly 1600 km s$^{-1}$ with channels separated by 3.3 km~s$^{-1}$, providing a velocity resolution of about 4 km~s$^{-1}$. These observations provide a $5\sigma$ {\sc Hi} column density sensitivity down to 1 $\times$ 10$^{19}$ cm$^{-2}$ per spectral channel. Further details of the observations and data reductions can be found in \cite{westmeier_gas_2011,westmeier_gas_2013}. We clip channels at the edges of the full velocity range (keeping channels with 14~km~s$^{-1} < v <$ 246 km~s$^{-1}$, around the systemic velocity of $\approx 131$km~s$^{-1}$) and clip pixels that do not have a flux above 6~mJy/beam in three consecutive channels for our analysis. 

\section{Search for Stellar Halo Substructure}
\label{sec:mf}

\begin{figure*}
    \centering
    \includegraphics[width=\linewidth]{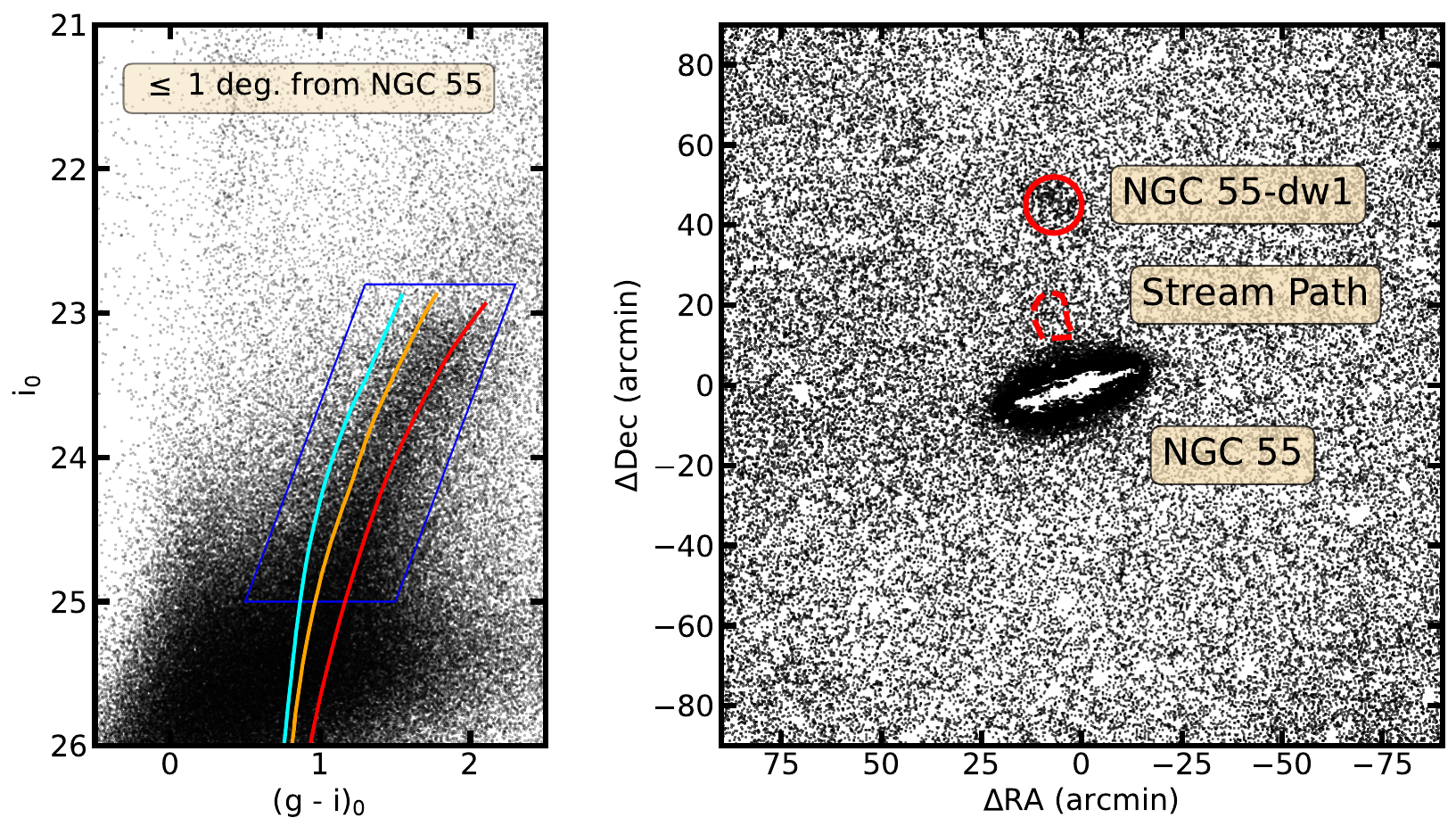}
    % \subfigure{\includegraphics[width=0.495\linewidth]{mf-3.2-1.8-annotated.png}}
    % \subfigure{\includegraphics[width=0.495\linewidth]{mf-10-1.8-annotated.png}}
    \caption{Left: A CMD of all sources that pass our star-galaxy separation criteria within one degree of NGC 55. We overplot three old, metal-poor (10 Gyr, [M/H] $=-2, -1.5, \text{ and} -1 $, respectively cyan, orange, and red) isochrones \citep[PARSEC,][]{bressan_parsec_2012,chen_improving_2014,pastorelli_constraining_2020} and a wide box indicating the location of the RGB at a distance of 2 Mpc ($\text{m}_{i,TRGB} \approx 22.8$). Right: The spatial distribution of all point sources within the RGB selection. NGC~55 appears in the center, although its inner regions are not recovered due to stellar crowding. A circle with a radius of 7~arcmin \citep[$2 r_h$ of dw1,][]{mcnanna_search_2024} is shown in red centered on dw1, and the selection we make for the stream is indicated with the dashed red line. A slight overdensity of sources is visible in the stream region. }
    \label{fig:sources}
\end{figure*}

The left panel of Figure \ref{fig:sources} shows a color-magnitude diagram (CMD) of our point sources within one degree of NGC 55. We include a wide selection to indicate the location of the red giant branch (RGB) to guide the eye, and overplot three 10 Gyr PARSEC isochrones \citep{bressan_parsec_2012,chen_improving_2014,pastorelli_constraining_2020} with [M/H]$=-2$, $-1.5$, and $-1$, shifted to 2 Mpc. The right panel shows a spatial plot of sources falling in the RGB selection. The main body of NGC~55 is clearly visible, and an overdensity of stars corresponding to dw1 is present to the north of the disk. A red circle with a radius of 7 arcmin \citep[twice the half-light radius of dw1, ][]{mcnanna_search_2024} is centered on the location of dw1. We also show the stream region identified in Section \ref{sec:mf} and Figure \ref{fig:mf-figure}. A slight stellar overdensity is visible in this region in the outer halo of NGC~55, although it is difficult to distinguish from background fluctuations in this plot. 

% We apply a matched-filter technique to better isolate coherent stellar substructure from background contamination. 

% A slight overdensity in this region is visible in the outer halo of NGC~55 in this plot, but a more robust search procedure is needed to make this feature more apparent.

% See \cite{rockosi_matched-filter_2002} for an in-depth description of this technique. 

% First applied in this context to find tidal tails around a Milky Way globular cluster \citep{rockosi_matched-filter_2002}, resolved-star matched filter detection algorithms have since been used to detect dwarf galaxies in mock LSST observations \citep{mutlu-pakdil_resolved_2021} and to search for signs of tidal disruption in the outskirts of nearby dwarf galaxies in deep, ground-based imaging \citep[e.g.,][]{sand_tidal_2012,mutlu-pakdil_deeper_2018,casey_deep_2025,prabhu_deep_2026}. In brief, this method maximizes the signal-to-noise ratio of data with a well-defined signal (in our case, a stellar population) that is embedded in well-behaved noise (here background and foreground sources). In brief, this method maximizes the signal-to-noise ratio of data with a well-defined signal (in our case, a stellar population) that is embedded in well-behaved noise (here background and foreground sources). We apply the following procedure to the full catalog of sources that passed our star-galaxy separation criteria. 

We apply a matched filter algorithm \citep{rockosi_matched-filter_2002} to search for stellar overdensities and substructure. In brief, this method maximizes the signal-to-noise ratio of data with a well-defined signal (in our case, a single stellar population) that is embedded in well-behaved noise (here background and foreground sources). To define our signal, we use PARSEC isochrones and luminosity functions with the Kroupa IMF \citep{kroupa_variation_2001} to create CMDs of two simulated stellar populations. They have ages of 3.2 and 10 Gyr, both with [M/H] = $-$1.8 dex, and are well-populated with $\sim$$8.5\times10^6$ stars. The old, metal-poor isochrone is a standard choice when searching for stellar halo substructure, and an additional younger isochrone is chosen to highlight the blue RGB as well as intermediate-aged asymptotic giant branch (AGB) populations. AGB stars extend to $\sim$$0.5$ mag above the TRGB in our filters and are effective in detecting potentially \emph{ex-situ} features in the outskirts of Local Volume dwarfs \citep{lee_star_2026}. We choose a [M/H] of $-1.8$ because it aligns with the blue edge of the prominent RGB feature in the CMD of sources near NGC~55 (Figure \ref{fig:sources}). We do not incorporate photometric error into our signal generation due to the wide CMD binning (described below), and we do not see any significant differences in the matched filter maps when the errors are included. 

% We tested perturbing the synthetic stars' magnitudes using the photometric errors to create our signal CMD, and found no appreciable differences in the resulting matched filter maps.}

We define the contaminant CMD using sources in an annulus spanning 1.5--3.5 degrees from the center of NGC~55. This region excludes the main body of NGC~55 and its nearby satellites while providing a sufficiently large area of empty sky to characterize foreground and background contamination. Both the signal and background CMDs are then binned in color-magnitude space (0.15 mag $\times$ 0.15 mag) to determine the relative density of sources in the CMD. We refer to the signal CMD as $ signal$ and the background CMD as $BG$ below. 

We then divide the full source catalog into $2\arcmin\times2\arcmin$ spatial bins. The calculation of expected sources in each bin aligned with the signal population is described in detail in Section 3 of \cite{rockosi_matched-filter_2002}.
% A CMD of the observed sources in each spatial bin is multiplied by the ratio of $signal$ to $BG$ across the CMD and summed. We then subtract the sum of the signal CMD and divide the result by the sum of the signal CMD squared divided by the noise CMD, which results in a number of expected counts in each on-sky bin aligned with the signal population $(N_{e})$. 
% This process is described in more detail in Section 3 of \cite{rockosi_matched-filter_2002}.
% The values are summed and then multiplied by the CMD bin area, giving a number of expected stars in each spatial bin $(N_{e})$. 
The distribution of expected stars across the field is smoothed with a Gaussian Kernel with a width $\sigma=$ 1.2 times the width of each bin, creating a smoothed map of expected counts $(N_{s})$. The median $(median)$ and standard deviation $(std)$ of $N_{s}$ are calculated using the \texttt{astropy.stats.sigma\_clipped\_stats} routine. We calculate the normalized signal as $(N_{s} - median)/std$ which gives the number of standard deviations each on-sky bin is above the median $(S)$. Contours are drawn around bins that have $S\geq4\sigma$ (Figure \ref{fig:mf-figure}).

% The sources in each bin are compared to the signal and contaminant CMDs, and then a number of expected sources aligned with the signal population is calculated for each on-sky bin. The contours are defined based on the distribution of resulting expected counts in each spatial bin: the vast majority of bins in the map contain predominately contamination and their expected counts fluctuate around zero, and bins containing an abundance of likely stars will have numbers $>0$. The on sky bins are smoothed with a Gaussian Kernel with a width 1.5 times the width of each bin, and contours are drawn around bins that have expected counts $\geq4\sigma$ above the background. 

\section{A Stream-Like Structure in the Halo of NGC 55}
\label{sec:results}

\begin{figure*}
    \centering
    \includegraphics[width=0.495\linewidth]{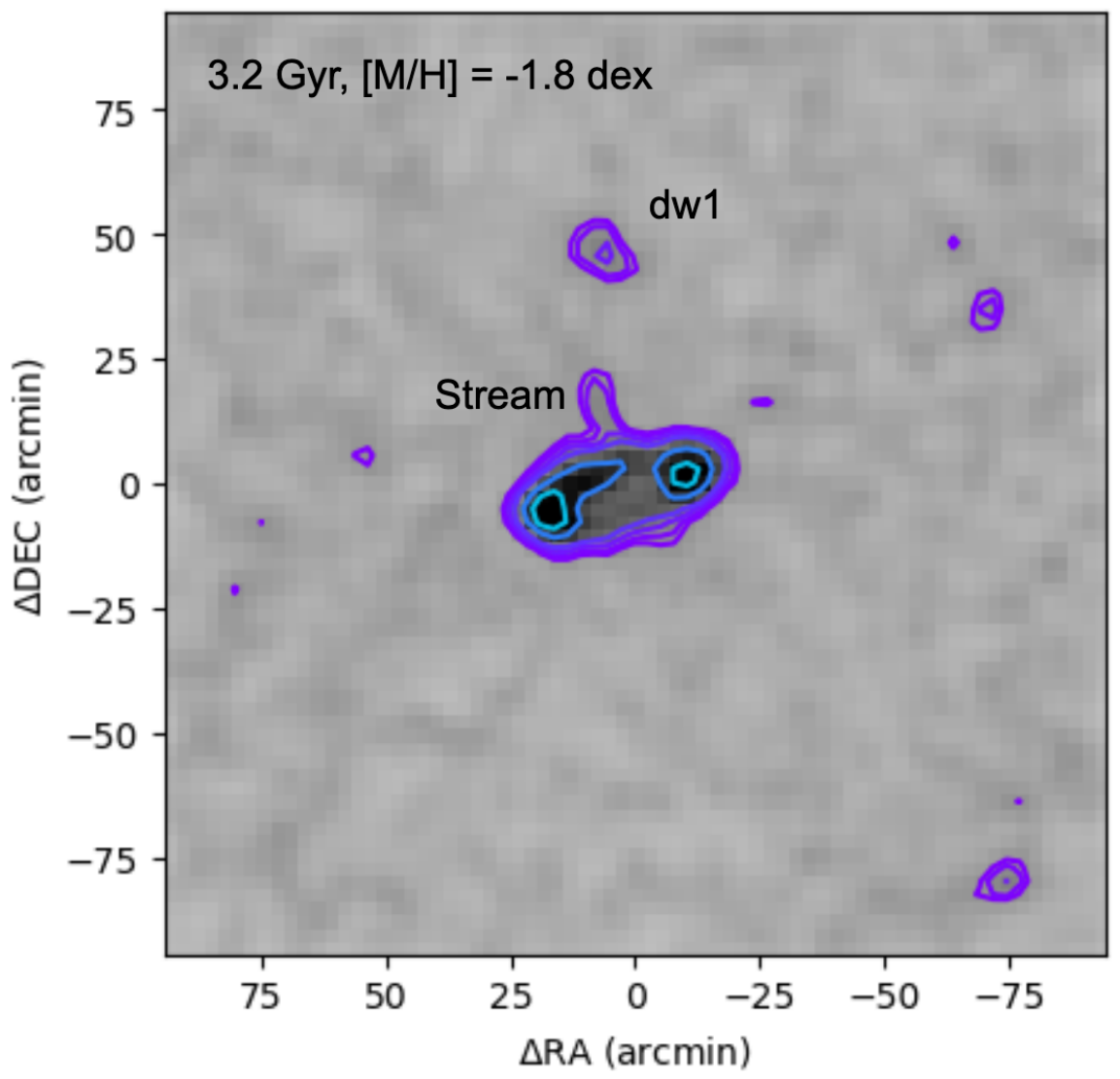}
    \includegraphics[width=0.495\linewidth]{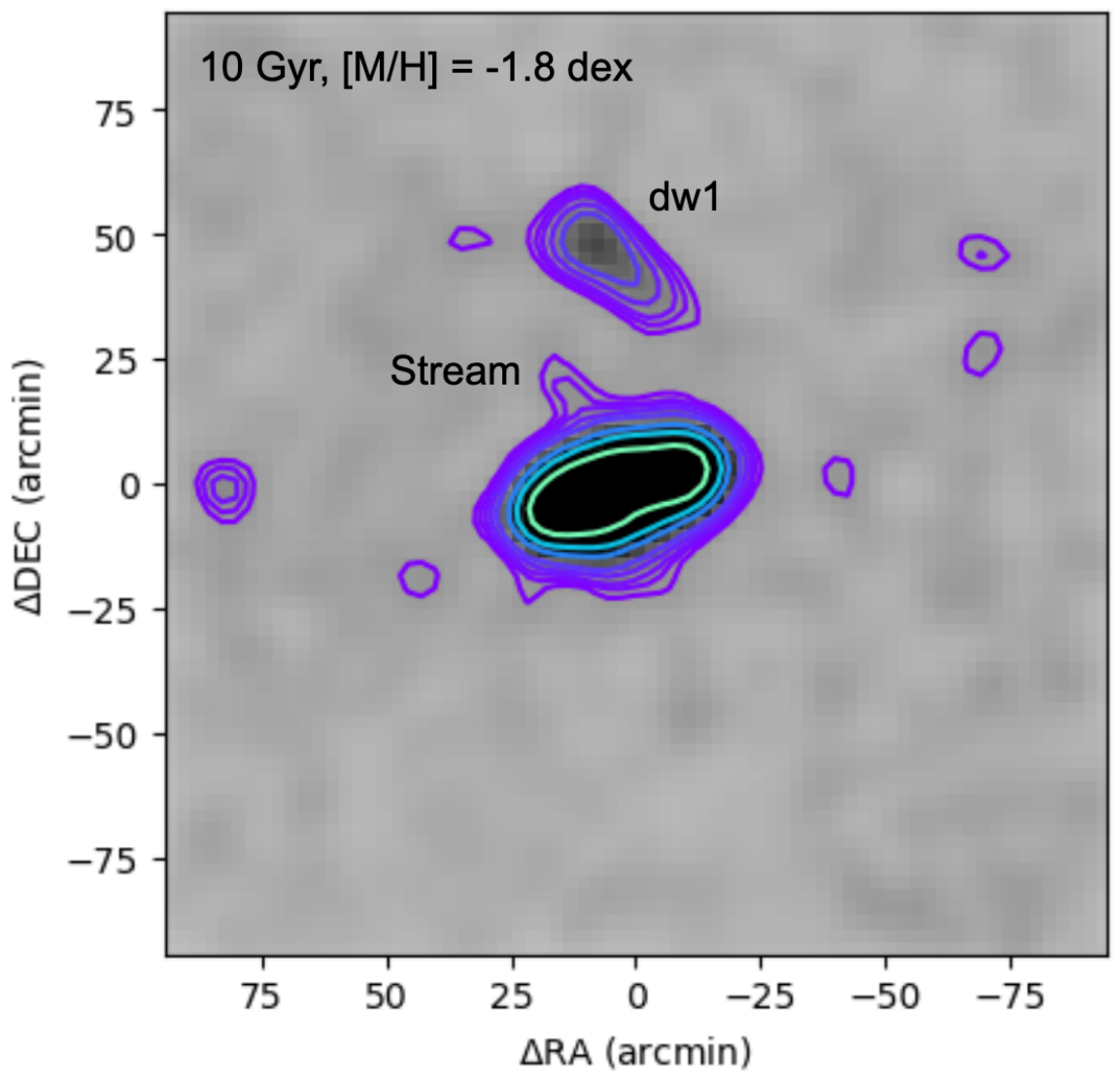}
    \caption{Two matched filter maps showing the stream in the halo of NGC 55, with contours showing $ (4,5,7,10,20,30,50,100)\sigma$ above the background. NGC~55 appears prominently in the center of the image, and the stream and dw1 are both labeled. Left: The matched filter map created using the 3.2 Gyr, [M/H] = -1.8 dex stellar population, which highlights bluer RGB stars and AGB stars. Right: The matched filter map made using the 10 Gyr, [M/H] = -1.8 dex population, which selects for slightly redder RGB stars and fewer AGB stars. In both panels, we recover dw1 with high significance, and we detect the stream at 5$\sigma$. The stream extends about 25 arcmin from the center of the galaxy, which corresponds to $\approx 14.5$ kpc at a distance of 2 Mpc. Overdensities in the background of these maps corresponds to masked foreground stars or compact background galaxy clusters. }
    \label{fig:mf-figure}
\end{figure*}

\begin{figure*}[tbh]
  \includegraphics[width=\linewidth]{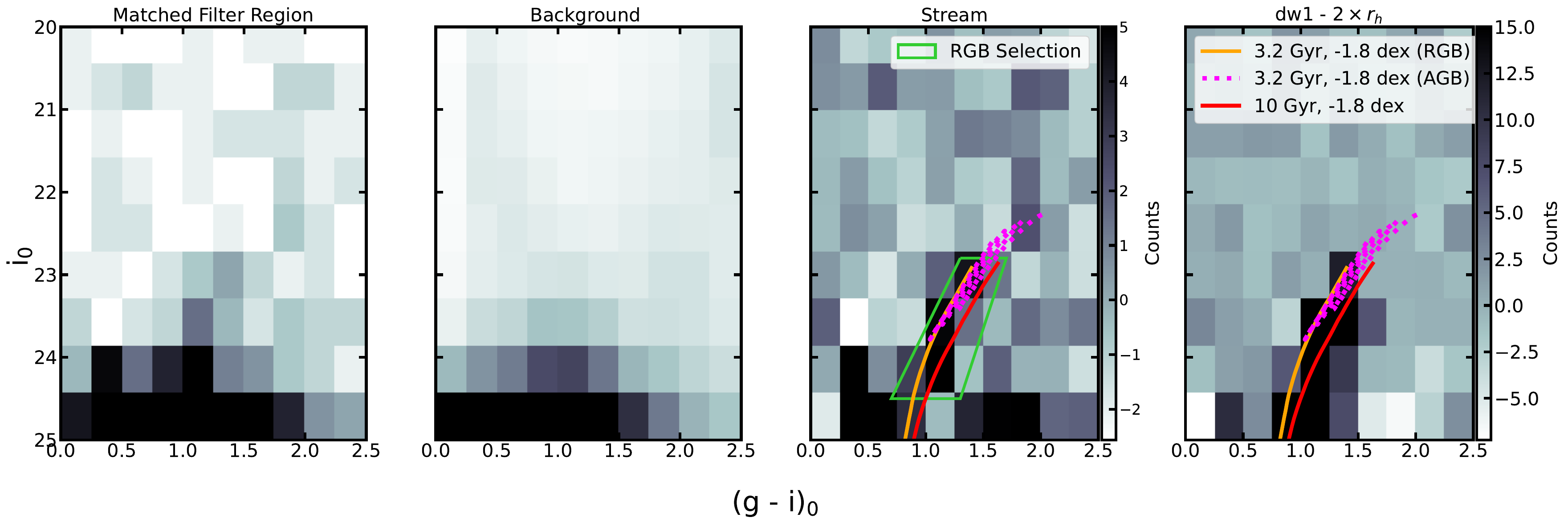}
  \caption{Hess diagrams of the stream selection (far left), the background region (center left), and the background-subtracted Hess diagrams of stars in the stream (center right) as well as within two half-light radii of the center of dw1 (far right). The panels in the far left show the area-weighted Hess diagrams before background subtraction. In the background subtracted panels, we show metal-poor ([M/H] = -1.8 dex) isochrones used to define the signals (red: 10 Gyr, orange: 3.2 Gyr RGBs, pink: 3.2 Gyr AGBs) shifted by the distance modulus of NGC~55 $(\mu = 26.61)$. These isochrones are consistent with the populated regions of color-magnitude space. The green box indicates the selection used to identify RGB stars in the analysis of the stream in Section \ref{sec:discussion}.}
  \label{fig:hess}
\end{figure*}

% \begin{figure}[tbh]
%   \includegraphics[width=\linewidth]{raw-cmd-stream-off.png}
%   \caption{Color-magnitude diagrams for our stream selection and an off-stream selection of the same area shifted by 30 $\approx 45$ arcmin northwest of the original selection. Isochrones are overlaid in the same manner as Figure \ref{fig:hess}. It is clear that there are many more stars aligned with the isochrones in the stream selection than in the of-stream region. }
%   \label{fig:raw-cmd}
% \end{figure}

\begin{figure*}[hbt]
  \includegraphics[width=\linewidth]{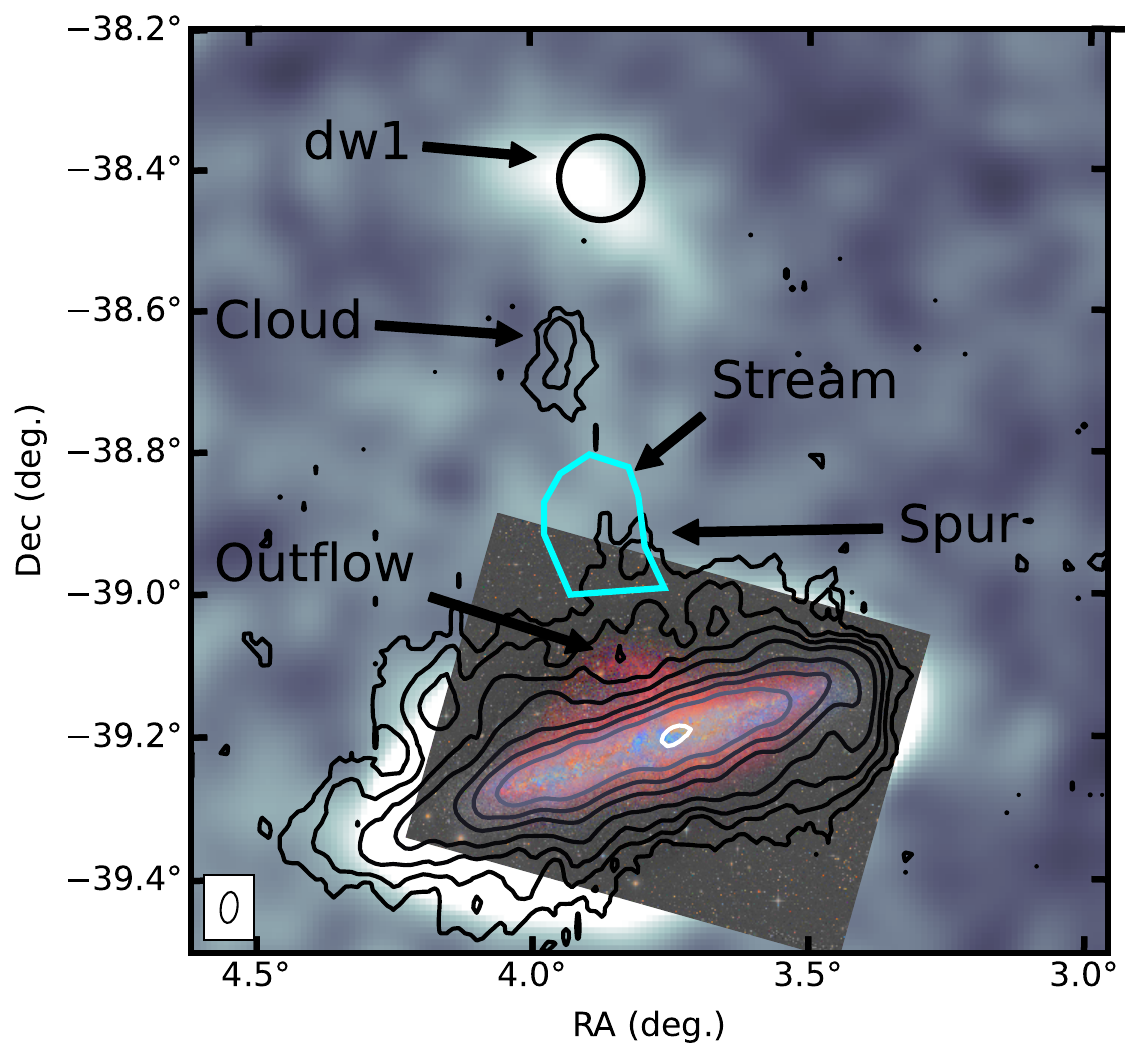}
  \caption{A KDE of the distribution of RGB stars. An image (image credit: Hanson, Dieterich, Zaytsev, Mazlin, Parker, Forman, \& Magill; https://www.hansonastronomy.com/ngc-55) showing the stellar disk and the H$\alpha$ fountain extending north of the disk is overlaid. We also show the {\sc Hi} moment 0 map (beam size of 158 $\times$ 84 arcsec$^2$, shown in the lower left). The contours show neutral hydrogen column densities of (0.1, 0.5, 1, 2, 5, 10, 20, 50)$\times \ 10^{20} \text{ cm}^{-2}$ \citep[as shown in ][]{westmeier_gas_2013}. A spur of gas is aligned with the stream and fountain, and all of these features point in the direction of an isolated cloud of {\sc Hi} and the nearby satellite NGC 55-dw1.}
  \label{fig:kde-hi}
\end{figure*}

\begin{figure*}[hbt]
    \centering
    \includegraphics[width=\linewidth]{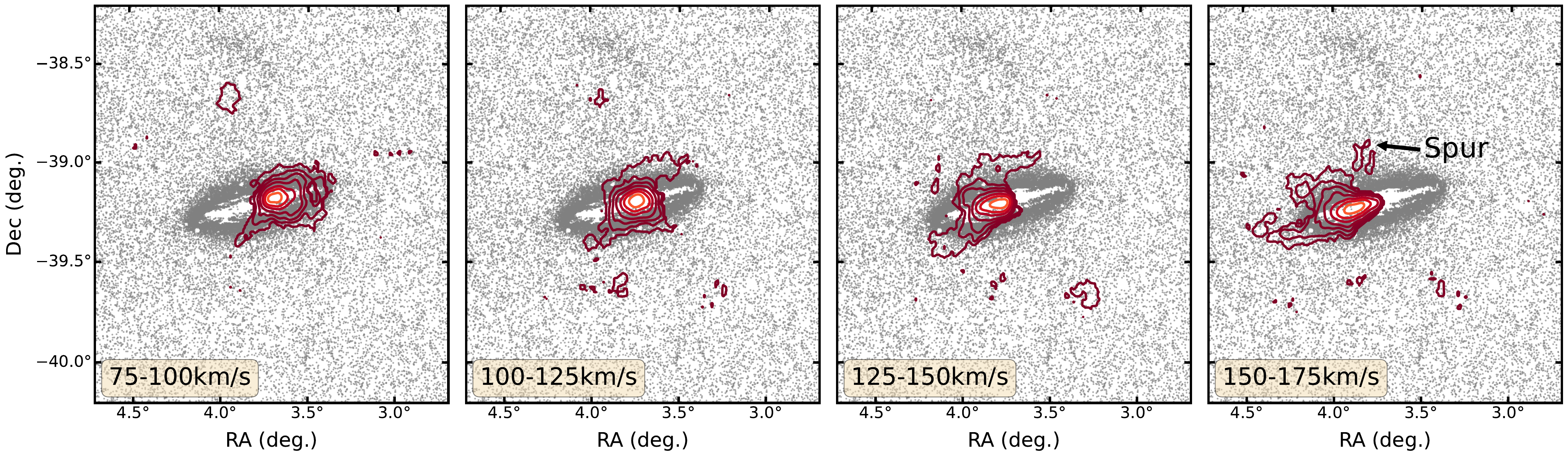}
    \caption{Four  {\sc Hi}  channel maps spanning 25 km/s, from 75 to 175 km/s plotted over the RGB source catalog. Contours are the same levels as shown in Figure \ref{fig:kde-hi}. The disk rotation of NGC~55 is visible from left to right. The spur can be seen in the rightmost panel on the center of the northern side of the galaxy. It appears in the same channel as the receding side of the disk, indicating that it is decoupled from the regular disk rotation.}
    \label{fig:channels}
\end{figure*}

Here, we present the matched-filter maps that reveal a stream-like structure in the halo of NGC~55, followed by a detailed analysis of its stellar populations. The left panel of Figure~\ref{fig:mf-figure} shows the matched-filter map centered on NGC~55 using the 3.2~Gyr stellar population as the signal.  The stream-like feature extends from the northern side of the galaxy in the direction of dw1. This feature extends 25 arcmin from NGC~55, or about 15 kpc in projection, from where the contours diverge from the outer disk. The right panel shows the same region, using the 10 Gyr stellar population as the signal. The stream is also recovered in this map, although with a slightly different apparent trajectory, which may reflect small-number statistics in the underlying stellar population. Its recovery at a consistent location using both signal populations supports the interpretation that the feature is a genuine stellar overdensity associated with the NGC~55 system. Both maps also show asymmetries on the southern side of the disk. Compact overdensities are associated with foreground stars (corresponding to ``holes" in the right panel of Figure \ref{fig:sources}) or clustered background galaxies and, unlike NGC~55, dw1, and the stream, are not consistently recovered in both matched-filter maps. 

The prominence of dw1 differs between the two matched-filter maps. In the map made using the younger stellar population (left panel of Figure \ref{fig:mf-figure}), dw1 is considerably less prominent than in the map made using the older population, indicating that its resolved stellar population is more strongly matched by the old, metal-poor isochrone. dw1 is asymmetric in both panels and very extended in the right panel of Figure \ref{fig:mf-figure} \citep[as reported in ][]{mcnanna_search_2024}. These features may indicate that dw1 is undergoing tidal disruption; we explore this possibility further in Section \ref{subsec:dynamics}.

We select candidate stream stars by tracing the contours shown in the left panel where the feature appears more prominent and spatially extended. 
The vertices of the resulting polygon, transformed from offsets relative to NGC~55 into celestial coordinates, are provided in Appendix \ref{appendix:stream}. We detect 215 sources in this region that fall within the RGB selection shown in Figure \ref{fig:sources}. Comparing this to $10^4$ regions randomly sampled within the background annulus (see Section \ref{sec:mf}) of the same area as the stream ($\mu=120$, $\sigma=21$; see Appendix \ref{appendix:bg-char}), this corresponds to a $4.52\sigma$ excess, consistent with our matched-filter detection. The probability that our structure originates from a random background fluctuation is $5 \times 10^{-6}$, leading us to conclude this detection is highly significant.

As an additional significance test, we construct Hess diagrams of our stream feature and five halo selections at the same distance from NGC~55 as the stream region. The Hess diagrams are created by binning the sources of interest in color-magnitude space and subtracting an area-scaled background population defined by a circle with a radius of 45 arcmin centered 1.75 degrees southwest of NGC~55. We subtract the mean Hess diagram of the halo selections from the stream, and recover the prominent RGB seen in Figure \ref{fig:hess}. This indicates that the structure identified by the matched filter is significantly different than other parts of NGC~55's stellar halo. See Appendix \ref{appendix:bg-char} for more information.

Figure \ref{fig:hess} shows the Hess diagrams of sources within the stream selection and within the background region. The background-subtracted Hess diagram, with PARSEC isochrones overlaid, is shown in the center-right panel. A clear RGB population is recovered, along with a small excess of sources consistent with AGB stars, suggesting the possible presence of an intermediate-age population. We also find tentative evidence for younger populations ($\sim0.5-1$~Gyr) at \emph{g}$-$\emph{i}$\sim0.5$, although contamination at these faint magnitudes prevents a robust characterization of this component. See Section \ref{subsec:dynamics} for a discussion of possible formation scenarios for young stars associated with the stream. 

Figure \ref{fig:hess} also shows the Hess diagram of dw1, constructed using stars within a radius of 7~arcmin, corresponding to twice its half-light radius \citep{mcnanna_search_2024}. Both the stream and dw1 show RGB populations consistent with old, metal-poor stars at approximately the same distance. However, there are possible differences between their stellar populations. In particular, the stream Hess diagram shows a small excess of sources consistent with AGB stars along the pink isochrone, while no comparable feature is apparent in dw1. However, the stream contains relatively few resolved stars, making a direct comparison with the stellar population of dw1 difficult. Deeper, likely space-based, photometry will be required to robustly characterize and compare their stellar populations (see Section \ref{subsec:limits-future}).

Figure \ref{fig:kde-hi} shows a Gaussian kernel density estimate (KDE) of the on-sky distribution of RGB stars. The sources are binned in (3 $\times$ 3 arcmin), smoothed with a kernel 3.5 arcmin wide. An image highlighting H$\alpha$ emission of NGC~55's disk taken at the Stan Watson Observatory South in El Sauce, Chile is also shown (see Appendix \ref{appendix:rgb} for further details of these observations). A prominent H$\alpha$ filament (marked as an "outflow" in Figure \ref{fig:kde-hi}) is clearly visible in the northern side of the disk below the stream region. The {\sc Hi} moment zero map is displayed as black contours (column densities of (0.1, 0.5, 1, 2, 5, 10, 20, 50)$\times \ 10^{20} \text{ cm}^{-2}$). The extent of the H$\alpha$ emission relative to the N$_{\text{\sc Hi}} = 5\times 10^{20} \text{cm}^{-2}$ contour and its confinement to a small area classify this feature as a ``fountain" \citep{mcquinn_galactic_2019}. Outflows like these are often driven by recent and active star formation, and are sometimes observed to trigger star formation in the outskirts of galaxies \citep[e.g.][]{rao_stars_2025}. See Section \ref{subsec:dynamics} for a discussion of a potential connection between the fountain and the stellar stream.

Our stream selection from the matched filter map is shown in cyan. The fountain is aligned with our stellar stream and with a spatially coincident {\sc Hi} spur. All of these features point toward an isolated {\sc Hi} cloud along the same projected direction as dw1. We find no evidence for a stellar overdensity coincident with this cloud in the matched-filter maps. We select a region around the clump and an equal-area background region offset to the west by one degree, and we see $\approx40$ more counts in the cloud region than the background. There is no clear RGB sequence in the CMD of sources selected around this region. Most of these sources are clustered below $i = 24$, implying that they are background galaxies. To quantify the stellar mass limit implied by this non-detection, we inject a synthetic stellar population (10 Gyr, [M/H]=$-2$~dex, including the photometric errors) into the CMD of this region. We are able to recover an RGB signal in the Hess diagram if the mass of the population is above $\approx 5\times 10^4 \ M_\odot$. This effectively places an upper limit on the stellar mass of any population spatially coincident with the {\sc Hi} cloud.
%We find no evidence for a stellar overdensity coincident with this cloud in our matched-filter maps, nor any clear RGB signal in a CMD of a spatial selection of this region. 

Narrower channel maps ranging from 75-175~km~s$^{-1}$ are shown in Figure \ref{fig:channels}. From left to right, these panels depict the coherently rotating {\sc Hi} disk of NGC~55 as well as the {\sc Hi} clumps in its outskirts. These {\sc Hi} features are known and extensively analyzed in \cite{westmeier_gas_2013}. We focus on only one of the five clouds identified by \citet{westmeier_gas_2013}; this is due to the apparent alignment of this cloud with the stream and dw1. The spur of {\sc Hi} is kinematically distinct from the disk rotation in the rightmost panel, as it appears above the center of the galaxy in velocity channels that are otherwise populated by the receding side of the disk. This indicates that this spur has been disturbed from the regular rotation of this galaxy. We focus on this feature in particular as it is aligned with the stellar and gaseous features of interest, and is more extended than other prominences in the channel maps. For the first time, we are placing the {\sc Hi} in the context of NGC~55's interaction history as traced by resolved stars. We explore the potentially shared origins for these stellar and gaseous features in Section \ref{sec:discussion}. 

\section{Discussion}
\label{sec:discussion}

We begin by discussing this stream and associated {\sc Hi} features in the context of previous studies of NGC~55 (Section \ref{subsec:prev-stud}). We then explore three possible progenitor scenarios for this stellar stream (Sections \ref{subsec:ind-prog}, \ref{subsec:dynamics}). We also provide a comparison with NGC~300, another LMC-analog in the Sculptor Group with recently discovered stellar substructures in its halo \citep[Section \ref{subsec:compn300}, ][]{fielder_streams_2025}. We end with a discussion of the current limitations of this work and prospects for further constraining the merger history of NGC~55 (Section \ref{subsec:limits-future}).

\subsection{Interpreting the Stellar and Gas Features in the Context of Prior Studies of NGC~55}
\label{subsec:prev-stud}

To interpret this stream, we need to consider it in the context of the NGC~55 system. NGC~55 is a nearly edge-on \citep[\emph{i} of $\sim$$80$ deg.,][]{kiszkurno-koziej_stellar_1988} barred disk galaxy \citep{de_vaucouleurs_third_1991} with a substantial {\sc Hi} component \citep[{\sc Hi} mass of $\sim10^9 \text{ M}_\odot$,][]{westmeier_gas_2013}. It is known to host multiple stellar populations in its disk with a range of ages \citep[e.g.][]{tikhonov_thick_2005} and a negative stellar metallicity gradient \citep[e.g.][]{kudritzki_spectroscopic_2016,patrick_physical_2017}. It has long been hypothesized that NGC~55 recently experienced a close interaction or merger \citep{hummel_neutral_1986,puche_h_1991} based on asymmetries in the morphology and kinematics of the observed {\sc Hi} gas and stellar distribution. 

\citet{mouhcine_halos_2005} used HST/WFPC2 imaging north of NGC~55 and identified a diffuse population of old, metal-poor RGB stars ($\text{[M/H]}\approx-1.1$), providing evidence for an extended stellar halo. Their small fields lie much closer to the disk than the stream detected here and were not sensitive to a structure of its extent and location.

Using Subaru/Suprime-Cam resolved-star imaging, \citet{tanaka_structure_2011} identified an asymmetric thick disk, two diffuse substructures northeast of the disk, and a diffuse halo component, which they interpreted as signatures of a past accretion event. We recover similar asymmetries in the disk but also detect the more extended stream-like structure presented here. Its absence from their maps may reflect their smaller field of view and differences in matched-filter construction and background characterization.

The thickening of the stellar disk on the eastern side is mirrored in the {\sc Hi} morphology (Figure \ref{fig:kde-hi}). \cite{westmeier_gas_2013} suggested that a potential previous merger or interaction caused the spurs and clumps of gas seen in the outskirts of the galaxy. The spatial alignment of the stellar stream and dw1 with these {\sc Hi} features, together with their distinct kinematics relative to the disk (Figure \ref{fig:channels}), lends further support to an interaction-driven origin. Taken together, these observations motivate an \emph{ex-situ} origin for the stream. We consider three scenarios: 1) the stream results from a single accretion event unrelated to dw1 (Section \ref{subsec:ind-prog}), 2) the stream is an extended tidal feature of dw1 on a low-eccentricity orbit (Section \ref{subsec:dynamics}), or 3) the stream originated from dw1 on a high-eccentricity orbit that passes through the disk of NGC~55 (Section \ref{subsec:dynamics}). 

\subsection{Independent Progenitor}
\label{subsec:ind-prog}

\begin{figure*}
  \includegraphics[width=\linewidth]{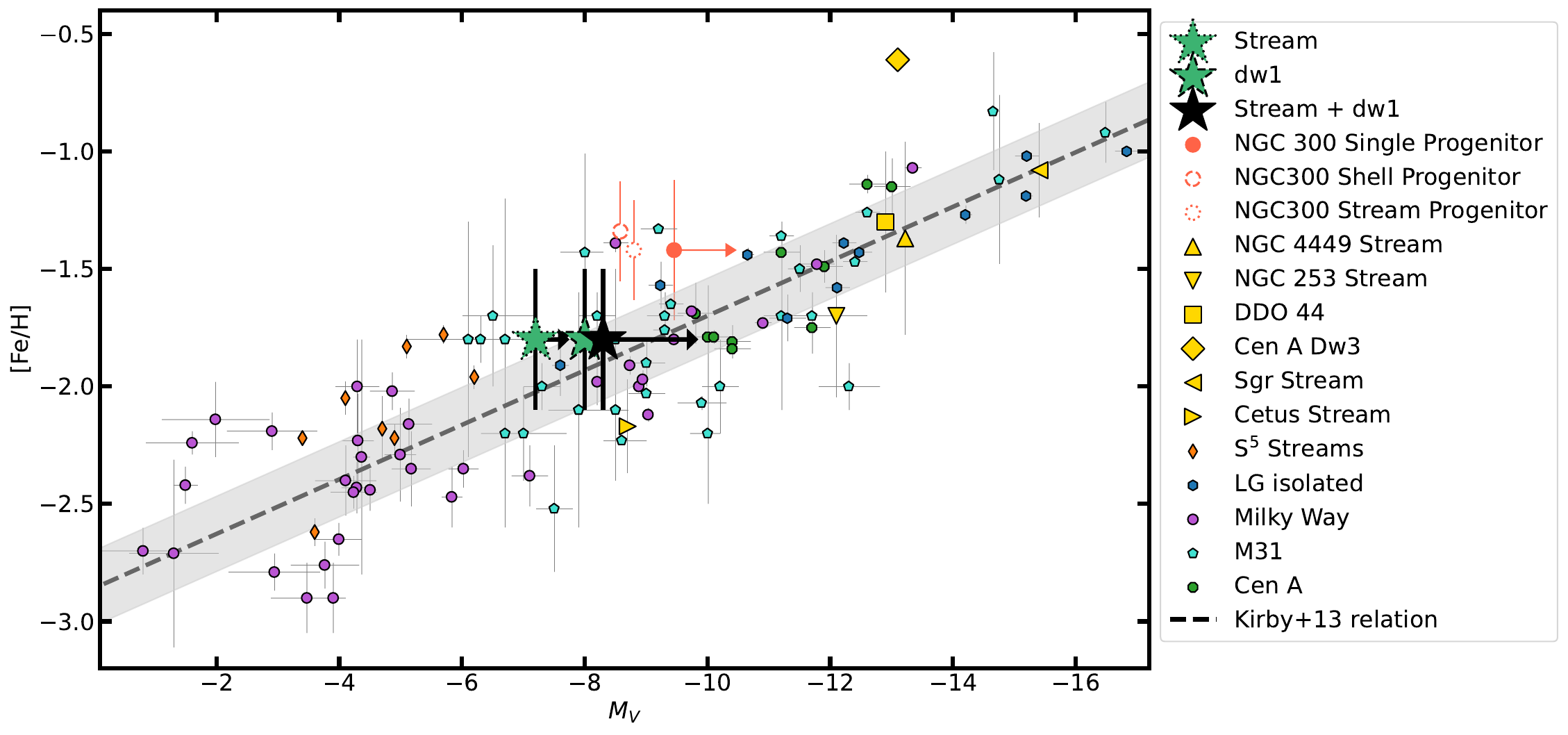}
  \caption{The local luminosity-metallicity plane of satellites and streams, adapted from \cite{collins_observational_2022} and \cite{fielder_streams_2025}. Points labeled with massive hosts correspond to their satellites. The best-fit relation and $1\sigma$ scatter from \cite{kirby_universal_2013} are overlaid. We show our stream, dw1, and combined stream+dw1 points in this relation. Both the stream and dw1 are consistent with the relation within their uncertainties, and their combined luminosity lies on top of the local relation. }
  \label{fig:massmet}
\end{figure*}

To test whether the stream could result from a single accretion event unrelated to dw1, we first estimate the luminosity of the stellar population within the stream selection shown in the third panel of Figure \ref{fig:hess} and in cyan in Figure \ref{fig:kde-hi}. We adopt a metallicity of [M/H] $=-1.8$ and an age of 10 Gyr for the stars in the stream, as the RGB is the most unambiguous feature in the Hess diagram. We note that the adopted metallicity is not a direct measurement, but is simply an illustrative value chosen with wide uncertainty. We sum the flux from all stars above the faint edge of our RGB cut, and then subtract the light from background sources using the same background selection as described in Section \ref{sec:results}. We also tested subtracting the total light contained in the stellar halo selections (Section \ref{sec:mf} and Appendix \ref{appendix:bg-char}) and found that this had no significant effect on our results. Assuming a Kroupa IMF \citep{kroupa_variation_2001}, we estimate that 88\% of the \emph{g}-band and 75\% of the \emph{i}-band luminosity of this stellar population falls below our RGB selection. We use this fraction to estimate the total flux of the underlying population. Using this corrected flux, we calculate its absolute \emph{V}-band magnitude $M_V$ using the transformations from DES DR2 \citep{abbott_dark_2021} adopted for the \emph{g}- and \emph{i}-bands, $M_V = 0.644M_g + 0.356M_i - 0.03$. We find that $M_V = -7.5 \pm 0.3$, with the uncertainty calculated using standard Gaussian propagation of uncertainties on the above formula for $M_V$. We consider this a lower limit, as it is likely that stars originating in the progenitor are projected in front of or behind the disk of NGC~55. We place this possible accreted satellite on the local luminosity-metallicity relation (\citealt{kirby_universal_2013}, see also \citealt{riley_auriga_2026}) in Figure \ref{fig:massmet} (dotted black/green star) adopting a metallicity uncertainty of 0.3 dex to account for systematic model uncertainties. Varying the metallicity of the population within the adopted $\pm0.3$ dex range does not affect the derived luminosity outside our uncertainties. We note that the relation presented in \cite{kirby_universal_2013} was derived using spectroscopic [Fe/H] measurements, as opposed to the photometric [M/H] inferred from isochrones that we use as a proxy. We also show various streams and Local Group galaxies \citep[adapted from][]{collins_observational_2022,fielder_streams_2025}. The stream is consistent with the line within the wide uncertainties in metallicity. 
% Importantly, disrupted systems are predicted to lie above the local relation \citep{riley_auriga_2026}. This may suggest that the stars in the stream compose the majority of the independent progenitor if they are indeed old and metal poor as modeled here.

We estimate the stellar mass of the detected stream using Equation~8 of \citet{taylor_galaxy_2011}, which relates \emph{g}$-$\emph{i} color and \emph{i}-band luminosity to stellar mass. We find $\log(M_*/M_\odot)=5\pm0.3$. The uncertainty includes the uncertainties in the measured color and \emph{i}-band luminosity, together with the intrinsic 0.1~dex scatter in the stellar mass-to-light relation \citep{taylor_galaxy_2011}. Varying the metallicity of the assumed population does not alter the derived stellar mass beyond our quoted uncertainties. As with the luminosity estimate above, we consider this a lower limit on the stellar mass of the progenitor. Additionally, if the candidate AGB stars in the stream trace an intermediate-age population, their presence may favor a progenitor more massive than $\sim10^{5}M_\odot$, as galaxies at and below this mass are typically dominated by ancient stellar populations. We next consider scenarios in which dw1 is the progenitor of the stream.

%We follow Equation 8 in \cite{taylor_galaxy_2011} to estimate the underlying stellar mass of the observed population. We find a total stellar mass of $\text{log}(M_*/M_\odot) = 5 \pm 0.25 $ in the stream. The error is calculated by adding the spread in color, \emph{i}-band magnitude, and intrinsic uncertainty of 0.1 dex \citep{taylor_galaxy_2011} in quadrature. We similarly consider this to be a lower limit of the stellar mass of the progenitor system for the reasons stated in the previous paragraph. Additionally, the possible AGB stars we observe in the stream suggest that the progenitor of this structure had a stellar mass larger than $\approx 10^{5} \ M_\odot$. We now consider scenarios where dw1 is the stream's progenitor.

% \subsection{dw1 on a Low-Eccentricity Orbit}
% \label{subsec:circ-dw1}
\subsection{Possible Association with dw1}
\label{subsec:dynamics}
% bolding the title since the whole section is different

The alignment of our stream candidate, dw1, an {\sc Hi} spur, and an isolated {\sc Hi} cloud associated with NGC~55 (Figure \ref{fig:kde-hi}) motivates us to investigate scenarios in which these structures share a common origin. We show dw1 in Figure \ref{fig:massmet}, adopting the measured $M_V$ $(-8^{+0.5}_{-0.3})$ and [Fe/H] $(-1.8$ dex) from \cite{mcnanna_search_2024}. We adopt a systematic uncertainty of 0.3 dex on the metallicity of dw1; \cite{mcnanna_search_2024} report no error on the [Fe/H] due to the low number of bright member stars. Within uncertainties, dw1 alone is consistent with the local luminosity-metallicity relationship. We combine the luminosity of the stars we detect in the stream with that of dw1, and find that the combined progenitor system has an $M_V = -8.5^{+0.2}_{-0.3}$. This gives a upper magnitude limit of $M_V \leq -8.3$. Adding the lower limit inferred stellar mass of the stream to the lower value reported for dw1 \citep[$1.42^{+0.45}_{-0.56}\times10^{5}$][]{mcnanna_search_2024} together, we find a stellar mass limit of $\log(M_*/M_\odot)\geq5.2$. The combined progenitor again is consistent with the local luminosity-metallicity relation within the large uncertainties \citep{kirby_universal_2013}, motivating further analysis. Below, we describe interaction scenarios where dw1 and the stream share a progenitor, which could produce the observed morphology of the gas, stream, and dw1. We then detail the simple dynamical models used to test these hypotheses. 

First discovered by \cite{mcnanna_search_2024}, dw1 is the most diffuse galaxy observed at its luminosity to date. With a low luminosity and large physical size \citep[$M_V \approx -8 \text{ and } r_{1/2} \approx 2.2 \text{ kpc}$, equating to a surface brightness of $\mu_g \approx 32.3 \text{ mag arcsec}^{-2}$,][]{mcnanna_search_2024}, dw1 falls squarely in the regime of ultra-diffuse galaxies \citep[UDGs, $\mu_g > 24.0 \text{ mag arcsec}^{-2}$ and $r_{1/2} > 1.5 \text{ kpc,}$][]{van_nest_whats_2022,li_beyond_2023}. The formation pathway(s) of ultra-diffuse galaxies are largely unknown \citep[e.g.][]{buzzo_extended_2026}, but one of the proposed mechanisms is through tidal interactions with a more massive host. UDGs in galaxy groups have been observed to possess tidal features \citep{bennet_evidence_2018}, and recently \cite{smercina_star_2025} characterized nearby UDG F8D1, a faint \citep[$M_V \approx -14$][]{caldwell_dwarf_1998} member of the M81 group with an extremely long \citep[$\approx 60$ kpc,][]{zemaitis_tale_2023} tidal tail. F8D1 and its host may provide an interesting comparison to NGC~55 and dw1, in which tidal interactions 
% with a relatively low-mass host 
could contribute to the extreme diffuseness of the satellite. 

% removed refernces to the bar structure
% Furthermore, NGC~2976 \citep{spekkens_modeling_2007} and NGC~55 are both barred, a signature of recent interactions in systems of their stellar mass. 

% Should future works confirm that NGC~2976 is the host to F8D1, NGC~55 and dw1 may be an analog to this other pair of galaxies, where tidal interactions have driven the barred kinematics of the host and diffuse morphology of the satellite.

% Radial mergers between galaxies often produce shell-like features,  We begin by discussing other studies of more massive galaxies that found shells in host galaxy halos, and connect these works to our observations and model of dw1 on a high-eccentricity orbit. 

Alternatively, radial mergers can produce shell- or umbrella-like stellar structures that may resemble diffuse satellites in projection. We therefore consider whether dw1 could represent a shell structure rather than an intact satellite. For instance, the MW-mass spiral galaxy M64 likely experienced a radial minor merger that produced a remnant consisting of a shell connected to the host by a diffuse stream-like feature \citep[see Figure 2 in][]{smercina_origins_2023}. \cite{martinez-delgado_giant_2023} found an massive ``umbrella" feature in the halo of NGC~922. They investigated this interaction history with N-body simulations, and found that a radial, off-center merger scenario with a satellite that has since been fully disrupted can reproduce the plume and stream-like features. 
The complex {\sc Hi} morphology of NGC~922 also favors a more complicated interaction history than a simple ``drop-through" encounter with the nearby satellite PGC~3080368 \citep{wong_ngc922_2006,elagali_h_2018}.
%These findings are in agreement with \cite{elagali_h_2018}, who show that NGC~922 must have experienced a more complicated merger history than the simple ``drop-through" merger scenario with nearby satellite galaxy PGC~3080368 \citep{wong_ngc922_2006} based on a more complicated and disordered {\sc Hi} morphology. 
NGC~55 may represent a lower-mass analog of these systems, in which dw1 is part of a more extended shell- or plume-like structure whose lowest-surface-brightness components remain undetected.
%where dw1 is actually an umbrella/plume like structure and we do not detect the full extent of the disrupted stellar populations due their low surface brightness. 
This radial merger scenario would provide a natural explanation for the disturbed {\sc Hi} kinematics, isolated {\sc Hi} in the halo, and the spur of gas aligned with the stream. 

\begin{deluxetable}{ccc}[bht]
\tablecaption{
    \textnormal{Properties of the potentials and distribution functions used to produce our toy dynamical models}
    \label{Tab:sim-props}
}
\tablecolumns{3}
\setlength{\extrarowheight}{4pt}
\tablewidth{\linewidth}
\tabletypesize{\small}
\tablehead{
\colhead{Property (Units)} & 
\multicolumn{2}{c}{Value}
}

\startdata
\hline
\multicolumn{3}{c}{\textbf{Potential}} \\ 
% \cmidrule(r){1-2}
 % & \textbf{Potential}  &  \\
% $M \ (M_\odot)$ & $2\times10^{10}$ & $2\times10^{10}$ \\
$M \ (M_\odot)$ & \multicolumn{2}{c}{$2\times10^{10}$}  \\
$R \ (\text{kpc})$ & \multicolumn{2}{c}{15}  \\
\hline
\multicolumn{3}{c}{\textbf{NGC~55}} \\
$M_* \ (M_\odot)$ & \multicolumn{2}{c}{$3\times10^{9}$}  \\
$R_* \ (\text{kpc})$ & \multicolumn{2}{c}{1.8}\\
$h_* \ (\text{kpc})$  & \multicolumn{2}{c}{0.5} \\
% $\Sigma_{0,*} \ (M_\odot\text{kpc}^{-2})$ & \multicolumn{2}{c}{1000} \\
$\sigma_{r,0,*} \ (\text{kms}^{-1})$ & \multicolumn{2}{c}{25} \\
$R_{\sigma,r,*} \ (\text{kpc})$ & \multicolumn{2}{c}{3}\\
$\sigma_{z,0,*} \ (\text{kms}^{-1})$ & \multicolumn{2}{c}{20} \\
$R_{\sigma,z,*} \ (\text{kpc})$ & \multicolumn{2}{c}{3} \\
$M_{\text{{\sc Hi}}} \ (M_\odot)$ & \multicolumn{2}{c}{$2\times10^{9}$} \\
$R_{\text{{\sc Hi}}} \ (\text{kpc})$ & \multicolumn{2}{c}{3}  \\
$h_{\text{{\sc Hi}}} \ (\text{kpc})$  & \multicolumn{2}{c}{1.5} \\
% $\Sigma_{0,\text{{\sc Hi}}} \ (M_\odot\text{kpc}^{-2})$ & \multicolumn{2}{c}{1000} \\
$\sigma_{r,0,\text{{\sc Hi}}} \ (\text{kms}^{-1})$ & \multicolumn{2}{c}{15} \\
$R_{\sigma,r,\text{{\sc Hi}}} \ (\text{kpc})$ & \multicolumn{2}{c}{5} \\
$\sigma_{z,0,\text{{\sc Hi}}} \ (\text{kms}^{-1})$ & \multicolumn{2}{c}{10} \\
$R_{\sigma,z,\text{{\sc Hi}}} \ (\text{kpc})$ & \multicolumn{2}{c}{5} \\
\hline
\multicolumn{3}{c}{\textbf{Shared Progenitor}} \\
$M_{DM} \ (M_\odot)$ & \multicolumn{2}{c}{$1\times10^{8}$} \\
$R_{DM} \ (\text{kpc})$ & \multicolumn{2}{c}{2.5} \\
$M_* \ (M_\odot)$ & \multicolumn{2}{c}{$1\times10^{6}$} \\
$R_* \ (\text{kpc})$ & \multicolumn{2}{c}{1.5} \\
$M_{\text{{\sc Hi}}} \ (M_\odot)$ & \multicolumn{2}{c}{$1\times10^{6}$} \\
$R_{\text{{\sc Hi}}} \ (\text{kpc})$ & \multicolumn{2}{c}{1.5}  \\
\hline
\multicolumn{3}{c}{\textbf{Progenitor Initial Conditions}} \\
 & Low $e$ & High $e$ \\
$(x_i,y_i,z_i)$ (kpc) & $(0, 5, 35)$ & $(0, -5, -30)$ \\
$(v_{x,i},v_{y,i},v_{z,i})$ (kms$^{-1}$) & $(-1, -35, 0)$ & $(1, 0, 15)$ \\
\hline
\enddata
\end{deluxetable}

% as a gas-rich spiral galaxy ($M_* = 3\times10^9 M_\odot$, \citealt{medoff_delve-deep_2024}, and $M_{\text{{\sc Hi}}} = 2\times10^9 \ M_\odot$, \citealt{westmeier_gas_2013}, see Table \ref{Tab:sim-props})

% To test if the stream is composed of stars associated with an interaction between NGC~55 and dw1, we produce two simple dynamical models of NGC~55 and a gas-rich dwarf spheroidal galaxy. We evolve them using different initial conditions to produce different orbital geometries, in an attempt to qualitatively reproduce the stellar and gaseous morphology in Figures \ref{fig:mf-figure} and \ref{fig:kde-hi}. We use the AGAMA code \citep{vasiliev_agama_2019} and GADGET-4 \citep{springel_simulating_2021} with further details given in this subsection. 

% To investigate the potentially shared origin of dw1's ultra diffuse nature and the halo features observed in NGC~55, 

To test if the stream is composed of stars associated with an interaction between NGC~55 and dw1, we construct two simple dynamical models of NGC~55 and a gas-rich dwarf galaxy. We adopt different initial conditions that place the satellite on low-eccentricity and high-eccentricity orbits, and examine whether either scenario can qualitatively reproduce the stellar and gaseous morphology shown in Figures \ref{fig:mf-figure} and \ref{fig:kde-hi}. We use AGAMA \citep{vasiliev_agama_2019} and GADGET-4 \citep{springel_simulating_2021} to model the past interaction between this pair of galaxies. AGAMA is a software library that provides a broad range of dynamical computational methods; we use it to define potentials from which we sample test particles to establish initial conditions for our toy dynamical models. These particles are evolved in time using GADGET-4 \citep{springel_simulating_2021}, a hydrodynamics code that evolves collisionless particles with traditional N-body methods and collisional fluids through smooth-particle hydrodynamics (SPH). These two codes are regularly used together to model interacting galaxies \citep[e.g.][]{weerasooriya_dancing_2026} as we do here.

We define the stellar potential of NGC~55 as a disk with a total mass of $M_* = 3\times10^9 M_\odot$ \citep{mcconnachie_observed_2012, dooley_predicted_2017}, a scale radius of 3~kpc, and a scale height of 0.5~kpc. These dimensions are chosen to approximate measurements of the thick and thin disk \citep{tikhonov_thick_2005,seth_study_2005,tanaka_structure_2011}. We sample 100000 star particles from this potential. The gaseous component of NGC~55 is also modeled as a disk, but with a total mass of $M_{\text{{\sc Hi}}} = 2\times10^9 \ M_\odot$, \citep{westmeier_gas_2013}, a scale radius of 5~kpc, and a scale height of 1.5~kpc. The adopted radial extent of the {\sc Hi} disk is motivated by the well-constrained {\sc Hi} mass-size relation for disk galaxies \citep{wang_new_2016}.
%These choices for the {\sc Hi} disk are based on the well-constrained {\sc Hi} mass-size relation \citep{wang_new_2016} of disk galaxies. 
This model produces a centrally peaked mass profile for the gas disk. Although more massive spiral galaxies often have a ``hole" in their {\sc Hi} density profiles \citep[e.g.][]{ott_vla-angst_2012}, \cite{westmeier_gas_2013} showed that NGC~55 has a peaked profile along its major and minor axes, as opposed to the ``double horn" profile expected in larger disk galaxies, supporting our adopted profile. 

We model dw1's progenitor dark matter halo with an NFW profile with halo mass of $M_{DM}=10^8 \ M_\odot$ and a scale radius of 2.5 kpc. We sample 100000 particles from this potential, and these are evolved in time alongside the stellar and gaseous particles. The stellar potential is modeled as a Plummer sphere, with a total stellar mass of $M_* = 10^6 M_\odot$ and a scale radius of 1.5 kpc. We sample 10000 star particles from this potential. We sample 10000 gas particles from the same spatial distribution function as the stellar component, also with a total mass of $M_{\text{\sc Hi}} = 10^6 M_\odot$. These parameters are chosen to approximately represent a Leo~P-like progenitor \citep{mcquinn_characterizing_2015}, a gas-rich dwarf galaxy with a stellar mass comparable to that inferred for dw1 \citep[Table 1;][]{mcnanna_search_2024}.
%We chose these characteristics to roughly mirror Leo P \citep{mcquinn_characterizing_2015}, a gas-rich dwarf galaxy with a similar stellar mass to dw1 \citep[Table 1,][]{mcnanna_search_2024}.

We evolve the NGC~55 and dw1 models in a fixed NFW potential with virial halo mass $M_{DM}=2\times10^{10} \ M_\odot $ \citep[NGC~55's halo mass,][]{westmeier_gas_2013} and a scale radius of 15 kpc. NGC~55 is set at the center of this potential. For the low-eccentricity case, we set the progenitor's initial position to $(0, 5, 35)$~kpc and initial velocity to $(-1, -35, 0)$ km s$^{-1}$. These initial conditions produce a fairly circular orbit around the disk of the host, where the orbital plane is perpendicular to the plane of the sky. In the high-eccentricity case, we set the satellite's initial position and velocity to $(0, -5, -30)$ kpc and $(1, 0, 15)$ km s$^{-1}$ (see Table \ref{Tab:sim-props}), keeping all other model parameters fixed. This setup produces a more radial orbit, in which the progenitor passes through NGC~55's disk while undergoing disruption. In both cases, the initial coordinates are chosen to produce the intended orbital geometries, approximately reproduce the observed stellar and gaseous features, and place the remnant near the present-day position of dw1. We choose not to model more complicated physics (i.e., the circumgalactic medium (CGM) of either galaxy, UV background radiation, multiphase gas, radiative transfer, star formation, etc.) for simplicity. These models are therefore simple illustrative tests of our combined-progenitor hypotheses, and should not be interpreted as realistic reconstructions of the true interaction history of NGC~55 and dw1.

\begin{figure*}
    \centering
    \includegraphics[width=\linewidth]{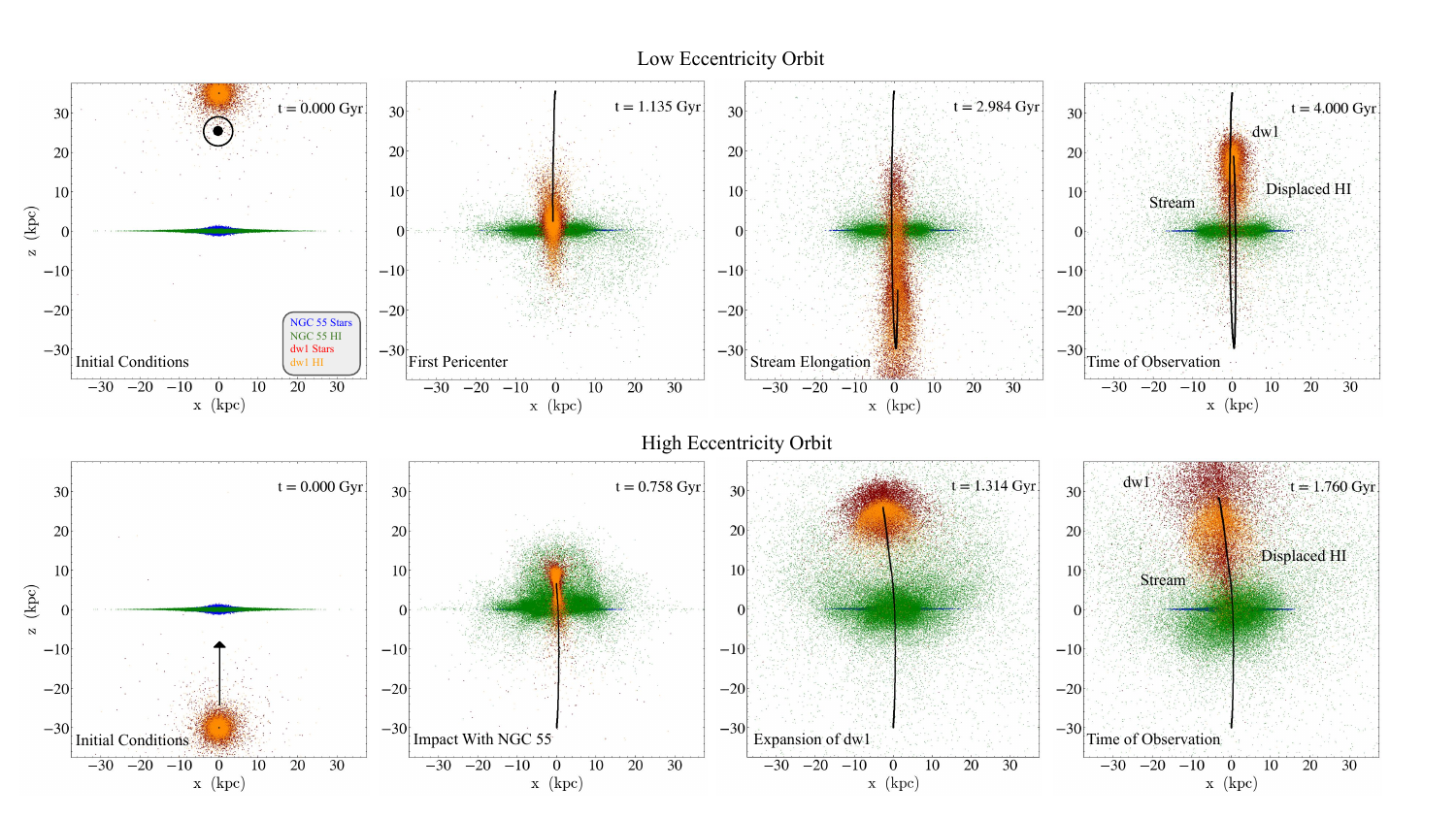}
    \caption{The results of our models to compare with the observed halo morphology (Figure \ref{fig:mf-figure} and \ref{fig:kde-hi}). Points are color coded by origin (blue:NGC~55 stars, red:dw1 progenitor stars, green:NGC~55 gas, orange:dw1 progenitor gas). The locations of the stream, disturbed {\sc Hi}, and dw1 are denoted in the rightmost panels. Orbit tracks are shown as the solid black line in each panel. Top: the four snapshots from our low eccentricity model. The initial velocity (out of the page) is indicated by the black dot in the left panel. Bottom: the four snapshots from our high eccentricity model. The initial velocity (toward the host) is shown by the black arrow. The stream morphology and the position of dw1 ($\approx 30$ kpc away from NGC~55 in projection) and its diffuse nature are qualitatively reproduced by both orbits.}
    \label{fig:dyn-model}
\end{figure*}

We present a series of snapshots of our models in Figure \ref{fig:dyn-model}, with the positive y-direction pointing into the page. The top row showcases the low-eccentricity orbit. We plot the direction of the initial velocity (out of the page) in the left panel. The final snapshot qualitatively reproduces the relative locations of the stream, dw1, and NGC~55, after the progenitor completes one full orbit in $\sim3.5-4$ Gyr. The stellar stream is co-spatial with the stripped gas from the satellite, which extends from the progenitor remnant toward the disk of NGC~55. The leading arm of the stream in the final snapshot wraps around NGC~55 toward the observer (out of the page), potentially lowering its projected surface density and explaining why we do not detect the stream over its full extent in Figure \ref{fig:mf-figure}. In contrast, the surviving dw1 remnant remains relatively concentrated, while only the higher-surface-density portion of the stream may be detectable above the background. This geometry also produces a distance gradient along the stream $(\approx 30 \text{ kpc between the stream and dw1} )$. In this model, the stream is composed entirely of stars stripped from the satellite. If dw1 and the stream share a common progenitor, the differences between their Hess diagrams (Figure \ref{fig:hess}) could reflect stellar population gradients within that progenitor. The surviving satellite remnant retains a small amount of {\sc Hi} associated with its stellar component. This gas could fall below the sensitivity of the existing {\sc Hi} observations \citep[$1\times 10^{19} \text{cm}^{-2}$,][ see Section \ref{subsec:hi} and Figure \ref{fig:kde-hi}]{westmeier_gas_2013}; wider and deeper {\sc Hi} observations are needed to confirm whether dw1 possesses a gaseous component. Additionally, this progenitor scenario results in the cloud and spur having different velocities along the line of sight, qualitatively matching what is observed about the velocities of these features in the channel maps (Figure \ref{fig:channels}). The actual velocity difference in this scenario depends on the inclination of the orbit with respect to our line-of-sight, which will require precise dynamical models to determine.  

The bottom row of Figure \ref{fig:dyn-model} presents four snapshots of the high-eccentricity orbital model, with the direction of the initial velocity vector indicated in the leftmost plot. The gaseous components of the progenitor and NGC~55 interact strongly in the second and third snapshots, producing a disturbed disk and displaced {\sc Hi} features qualitatively similar to those shown in Figure \ref{fig:kde-hi}. Stars primarily originating from the satellite populate the stream seen in the final panel, corresponding to the stream we observe in Figure \ref{fig:mf-figure}. The progenitor remnant corresponding to dw1 is also much more extended here than in the low-eccentricity model. The high-eccentricity merger perturbs the disk of NGC~55 more strongly than observed (Figures \ref{fig:kde-hi}, \ref{fig:channels}), suggesting that this particular realization is too disruptive to reproduce the present-day system. 

% To associate the {\sc Hi} with the satellite in this orbital configuration, the satellite progenitor must have had a lower {\sc Hi} mass to reproduce all of the observed characteristics of this system. Combined with the observed mass of the cloud \citep[$7.7\times10^6 \ M_\odot$][]{westmeier_gas_2013} this merger scenario is unlikely, unless this significant amount of {\sc Hi} existed in the progenitor without forming stars or more gas from NGC~55 was displaced than is apparent in our model. 

% composed of multiple ages/populations

In both models, the cloud of {\sc Hi} is disrupted from the host $\approx1-2$ Gyr before the time of observation. To associate the progenitor with the displaced {\sc Hi} cloud in our models, the gas must be as old or older than the predicted $\approx 1$ Gyr lifespan of isolated {\sc Hi} clouds \citep[][]{ivleva_merge_2024}, or the merger happened more recently than we model. In the case of a more recent merger, the shared progenitor likely continued forming stars until its encounter with NGC~55. Although we assumed an old and metal-poor population for both dw1 and the stream in our fiducial analysis, the ages of the stellar populations in the stream are unknown, and the age of dw1 is poorly constrained \citep[best-fit age of $6.5^{+4.3}_{-2.7}$ Gyr, with a broad posterior distribution][]{mcnanna_search_2024}. The stream Hess diagram also shows tentative features consistent with a younger ($\approx0.5$--1~Gyr) more metal-rich component, hinting that the star formation history of this structure is more complex than we assume. Space-based observations of this stream and dw1 may uncover younger populations and a common extended star formation history, lending credence to the shared progenitor formation pathway. 

% The ages of the stellar population in the stream are unknown, and the star formation history of dw1 is poorly constrained \citep[best-fit age of $6.5^{+4.3}_{-2.7}$ Gyr, with a broad posterior distribution][]{mcnanna_search_2024}.  

% It is important to note that the ages of the stellar population in the stream are ikely old but unknown, and the star formation history of dw1 is poorly constrained \citep[best-fit age of $6.5^{+4.3}_{-2.7}$ Gyr, with a broad posterior distribution][]{mcnanna_search_2024}. Although we adopt an old population for both dw1 and the stream in our fiducial analysis, the stream Hess diagram also shows tentative features consistent with a younger ($\sim0.5$--1~Gyr), more metal-rich component. 

% that the stream stars formed \emph{in situ} from gas associated with the observed outflow (Figure \ref{fig:kde-hi}), potentially through interactions between outflowing gas and the CGM of NGC~55.
%We assumed that the stars are old in both dw1 and the stream, but the Hess diagram of the stream can be fit by-eye with a more metal rich $\approx 0.5-1$ Gyr old population. The presence of an outflow (Figure \ref{fig:kde-hi}) associated with the other features we analyze mean that \emph{in-situ} star formation triggered by the outflow interacting with the circumgalactic medium (CGM) of NGC~55 could have created the stellar stream we observe. 

% The possibly young stars and isolated cloud additionally could share in origin related to the fountain (Figure \ref{fig:kde-hi}). Outflows can launch gas into the CGM of the host galaxy 

An alternative origin for the possibly young stars in the stream Hess diagram is through interactions between the outflow (Figure \ref{fig:kde-hi}) and CGM of NGC~55. Outflows from both active galactic nuclei and star formation are observed to form stars around larger galaxies after interacting with the CGM \citep{ong_signatures_2025,rao_stars_2025,rao_stars_2026}. It is possible that the stars we detect in the stream formed through this mechanism and are predominantly \emph{in-situ}. Observing a recent star formation history of the stream and an ancient one for dw1 would support this scenario. In more massive galaxies, the stars formed from outflows tend to be $\approx10-500$ Myr old \citep{crockett_triggered_2012,rao_stars_2025,rao_stars_2026}. 
% The stars formed in the inner filament of Centarus A's halo are $\lesssim 10$ Myr old \citep{crockett_triggered_2012}, and those formed through interactions between M82's outflow and CGM are $\approx$ 630 Myr old \citep{rao_stars_2025,rao_stars_2026}. 
Future theoretical work is necessary to predict the impacts and timescales of star formation triggered by outflows in the halos of low mass galaxies. It is worth noting that our work here provides possible origins for all of the analyzed structures, and the stellar and gaseous structures could originate from a combination of the scenarios we investigate. 

\subsection{Comparison to NGC~300}
\label{subsec:compn300}

We compare our results with those found by \cite{fielder_streams_2025} in the stellar halo of NGC~300. This system is well-suited for direct comparison given its close resemblance to NGC~55 as a host, and that as of now these are the only two galaxies in this mass range with resolved stellar halo substructure. Additionally, both of these dwarf spiral galaxies are observed in DELVE-DEEP \citep{drlica-wagner_decam_2022}, are at a distance of $\sim2$ Mpc, have a stellar mass similar to that of the LMC, are loose members of the Sculptor Group, and lie in relatively isolated environments. NGC~300 also had complimentary {\sc Hi} observations \citep[presented first in ][]{westmeier_gas_2011} that \cite{fielder_streams_2025} use to support their claim that previous accretion event(s) produced the structures they observe. 

While both of these dwarf galaxies likely experienced recent mergers or interactions that produced their stellar halo substructures, there are notable differences in their observed properties and inferred origins that we contrast here. The halo structures, both in number and morphology, are quite different between these two systems. NGC~300 hosts at least three distinct streams and shell-like features, the longest of which stretches 40 kpc from the galactic center. The potential origins are described in detail by \cite{fielder_streams_2025}, but it is unclear whether or not these structures originate from one or multiple progenitors. We detect only one distinct feature in NGC~55's halo. This hints that the merger experienced by NGC~55 was with a smaller satellite than that of NGC~300, reflected by the absolute magnitude of the various progenitor systems shown in Figure \ref{fig:massmet} (NGC~55 in black/green, NGC~300 in orange). 

Additionally, NGC~300's optical disk is almost perfectly exponential \citep[e.g.][]{de_vaucouleurs_southern_1962,carignan_surface_1985,bland-hawthorn_ngc_2005,vlajic_abundance_2009}, implying that the processes affecting the outskirts of this galaxy did not impact the central regions. This is reflected in the {\sc Hi} observations, which show only a slight disturbance in the direction of the longest stream \citep{westmeier_gas_2011,fielder_streams_2025}; the {\sc Hi} morphology in NGC~300 is considerably more regular than in NGC~55. In contrast, the stellar disk and {\sc Hi} distribution of NGC~55's are asymmetric, with a thickening of both components on the eastern side, which implies that the merger or interaction NGC~55 experienced had a much more dramatic effect on its main body than that of NGC~300. Since the inferred luminosities of each scenario imply that the progenitor(s) of the structures in NGC~300 were more massive, we hypothesize that the potential interaction experienced by NGC~55 was more recent. Detailed dynamical models are needed test this hypothesis. 

% This may indicate that the progenitor(s) responsible for NGC~300's halo substructure followed lower-eccentricity orbits than those inferred for NGC~55. Such orbits often produce more stream-like features than shells \citep[e.g.][]{karademir_outer_2019} and cause smaller dynamical perturbations to the inner regions of the host compared to a radial merger. This difference in disruption of the disk of the host galaxy is illustrated by our toy models (Figure \ref{fig:dyn-model}), where the {\sc Hi} disk of NGC~55 is significantly disrupted by the high-eccentricity merger, and is less disturbed by the low-eccentricity orbit. 

 % As previously discussed, there are also spurs and disturbed clouds of gas that are detached from the disk itself. This could imply that the merger experienced by NGC~55 happened more recently than that of NGC~300, where the gas would have had time to settle back into the disk. 

Together, these differences in stellar disks, gas morphology, and stellar halo substructure between these two systems paint complementary pictures of their merger histories, providing some of the first insight into the diversity of merger histories among isolated LMC-mass hosts. Future observations of these two systems will allow for the origins of these features to be fully constrained, and systematic searches for substructures around low-mass galaxies will allow for population-level studies of dwarf stellar halos to be performed for the first time. 

% The comparison with \cite{fielder_streams_2025} suggests progenitor orbits play a significant role in the resulting halo and disk morphology of the dwarf host galaxies. 

\subsection{Current Limitations and Future Prospects}
\label{subsec:limits-future}

% We have presented the discovery of a stream-like structure in the stellar halo of NGC~55. Its background-subtracted Hess diagram shows evidence of metal-poor stellar populations with ages $\gtrsim 3$ Gyr. It aligns with a kinematically decoupled spur of {\sc Hi}, a disrupted cloud of {\sc Hi}, and the ultra diffuse satellite dw1. We explored three potential progenitor scenarios for these features: the stream originating from a merger independent of dw1, the stream being an extended tail of dw1 from a merger on a low-eccentricity orbit, or these features coming from a high-eccentricity collision with dw1. %this feels conclusion-y, but i like having a single statemnt at the beginning of the limitations section to give an overview of what was done before going into the details of what might be wrong

% None of our investigations of the origins of these features are particularly constraining, nor do they rule one another out completely.

We have presented several scenarios that can qualitatively describe the gas and stellar features around NGC~55, although no scenario is conclusive. This is primarily due to the low number of stars that we can detect in this stellar stream feature ($\approx90$ RGB candidates and $\approx5$ AGB candidates above the mean background within our selection). Stellar halo features are inherently low surface brightnesses and observing the resolved stellar populations within these kinds of structures is even more difficult from the ground due to poor star-galaxy separation \citep{fadely_stargalaxy_2012}. Furthermore, matched filter searches like ours are more powerful for galaxies within the MW halo. At the distance we probe here, the observed populations are flooded by background galaxies at fainter magnitudes; even with very deep ground-based photometry, the contamination makes matched filter searches much less effective than for nearby systems. We need wide-area, space-based imaging to overcome this. 

% when the main sequence turn-off is detectable. Outside of 500 kpc, this relatively dim feature is flooded by faint, blue background galaxies

% the detectability of these structures are further suppressed when they are around dwarf galaxies.

% Additionally, dw1 was not discovered until 2024 \citep{mcnanna_search_2024}, a full eleven years after the observations of NGC~55 were published \citep{westmeier_gas_2013}. Deeper and wider follow up {\sc Hi} observations that cover dw1 could uncover additional {\sc Hi} associated with this satellite. 

Robustly modeling the dynamical history of this stellar halo feature and dw1 is not currently possible, due to both the number of stars in the detected features and the lack of kinematic measurements. Our toy models are merely illustrative of possible past interactions with dw1 needed to produce the stellar stream and gas morphology. In tandem with dynamical models, modeling the star-formation histories of the stream is necessary to determine the \emph{in-} versus \emph{ex-situ} nature of this feature. Our investigations into the stream assuming old and metal-poor populations favor an accretion origin for the stream, but the presence of an outflow and tentative evidence for younger stars in the Hess diagram mean that we can not rule out the \emph{in-situ} contribution. Future space-based observations of the stream and dw1 will provide the photometric depth and precision needed to better constrain their stellar populations, density structure along the stream, globular cluster populations, and star formation histories, helping to distinguish between these formation scenarios and determine the origin of the stream.

%required precisely model this system's star formation and interaction history to determine the origins of the stream.

Despite these limitations, the DELVE-DEEP observations provide the deepest wide-field imaging currently available for NGC~55. The survey reaches depths similar to that of the first year of the Rubin Observatory's LSST, where the $5\sigma$ point source detection limit is expected to reach \emph{g} $\approx 25$ and similar depths in the five other filters \citep{ivezic_lsst_2019}. Our matched-filter analysis enhances the contrast of stellar populations relative to foreground and background contaminants, while the Hess diagrams and spatial correspondence with dw1 and disturbed {\sc Hi} provide independent support for the reality of the stream-like structure and a possible connection to a recent interaction or merger.

\section{Conclusion}
\label{sec:conclusion}

In an ongoing effort to characterize the stellar halos of LMC- and SMC-mass galaxies in the DELVE-DEEP survey, we have discovered a stream-like structure in the stellar halo of NGC~55 using an implementation of the matched filter detection algorithm (Figure \ref{fig:mf-figure}, Section \ref{sec:results}). The structure extends $\sim15$ kpc in projection out of the disk and into the northern halo of NGC~55, and its background-subtracted Hess diagram shows evidence of metal-poor stellar populations with poorly constrained ages, although it is likely they are $\gtrsim 3$ Gyr old (Figure \ref{fig:hess}). This structure also aligns with a kinematically decoupled spur of {\sc Hi} (Figure \ref{fig:channels}; Sections \ref{sec:results}, \ref{subsec:prev-stud}) and an outflow of ionized gas (Figure \ref{fig:kde-hi}). Both of these features point toward a cloud of {\sc Hi} and the ultra-diffuse NGC~55 satellite dw1 (Section \ref{sec:mf}, \ref{sec:results}; Figures \ref{fig:mf-figure}, \ref{fig:kde-hi}). Assuming an old, metal-poor (10 Gyr, [M/H] = $-1.8$ dex) stellar population and correcting the observed flux for stars below our detection limits, we find that the stars contained in the stream structure have a total absolute magnitude $M_V \leq -7.2$ and stellar mass $\geq 10^5 M_\odot$ (Figure \ref{fig:massmet}). 

We investigated the possibility that this stream, the disrupted neutral hydrogen, and dw1 originate from a shared merger event. Combining the sources detected in our structure with those in dw1, we find that their shared progenitor had an absolute magnitude $M_V \leq -8.3$ and stellar mass $\geq 10^{5.2} M_\odot$, aligning well with the local luminosity-metallicity relation. We model the NGC~55 + dw1 system using AGAMA and evolve it with GADGET-4 using two different initial positions and velocities to produce two distinct progenitor orbits, one with a low eccentricity and one with a high eccentricity. The progenitor disrupts and leaves behind a diffuse concentration of stars, a stellar stream, and clumps of gas that mirror the observed morphology of NGC~55's halo and dw1 in both cases (Figure \ref{fig:dyn-model}), however the high-eccentricity model produces substantially more disruption in the disk of NGC~55 than is observed, disfavoring this particular realization. The spatially coincident outflow also leaves open the possibility that some stars formed \emph{in situ} through interactions between outflowing gas and the CGM of NGC~55. It is also possible that the observed structures originate from a combination of the scenarios investigated here. 

Our comparison to NGC~300, another LMC-mass dwarf galaxy recently found to host remnants of one or multiple merger(s) in its halo, gives the first look into the comparative population-level studies of dwarf stellar halos and merger histories that will be possible with upcoming observations. Within the DELVE-DEEP survey, there are two SMC-mass galaxies (Sextans~A and B), and the MADCASH survey observes seven more of these LMC/SMC-mass systems (four observed with Subaru/HSC and three with DECam). LSST will resolve stars in the outskirts of roughly 16 isolated (tidal index $\Theta < 0.5$) dwarf galaxies in this mass range \citep{mutlu-pakdil_resolved_2021}, three of which will be observed in the Roman High-Latitude Wide-Area Survey \citep[NGC~300, NGC~55, NGC~7793,][]{sanderson_near_2026} as well as the Euclid Wide survey \citep{euclid_collaboration_euclid_2022}. These complementary surveys will make it possible to detect dwarf stellar halos across a range of morphologies and environments. Combined with pencil-beam views from HST or JWST, it will be possible to fully characterize and understand the origins of structures like the one we present here, providing an unprecedented understanding of small-scale galaxy evolution and hierarchical growth. Our work provides a detailed observational look at a low-mass stellar halo and highlights the complexity of dwarf stellar halo assembly.

% NGC~247, NGC~7793, Sextans A, Sextans B,

% Combining the upcoming wide observations of NGC~55 with deep, .. and understand its origins,

% , and to trace their diverse formation pathways

% Our work supports the idea that LMC-mass dwarfs assemble their stellar halos in the way massive galaxies do: through the hierarchical accretion of satellites.

% \twocolumn
\appendix 
\section{Location of The Stream Selection}
\label{appendix:stream}

Table~\ref{Tab:appendix-coords} lists the right ascension and declination of the vertices defining our stream-selection polygon. The polygon was traced from the left matched-filter map in Figure~\ref{fig:mf-figure} and transformed from angular offsets relative to NGC~55 into celestial coordinates.
%In Table \ref{Tab:appendix-coords} we show the right ascension and declination for the polygon we use to do our stream analysis. This selection was traced from the left matched filter map of Figure \ref{fig:mf-figure} and converted from arcmin from NGC~55 to Ra. and Dec. 

\begin{deluxetable}{cc}[bht]
\tablecaption{
    \textnormal{Coordinates of the stream selection vertices from the matched filter map}
    \label{Tab:appendix-coords}
}
\tablecolumns{2}
\setlength{\extrarowheight}{4pt}
\tablewidth{\linewidth}
\tabletypesize{\small}
\tablehead{
\colhead{Right Ascension (Deg.)} & 
\colhead{Declination (Deg.}
}
\startdata
\hline
3.811 & -38.865 \\
3.829 & -38.825 \\
3.899 & -38.807 \\
3.951 & -38.834 \\
3.980 & -38.874 \\
3.980 & -38.919 \\   
3.934 & -39.004 \\
3.764 & -38.995 \\
3.799 & -38.936 \\
\hline
\enddata

\end{deluxetable}

\section{Significance of the Stream Detection}
\label{appendix:bg-char}

\begin{figure}
    \centering
    \includegraphics[width=\linewidth]{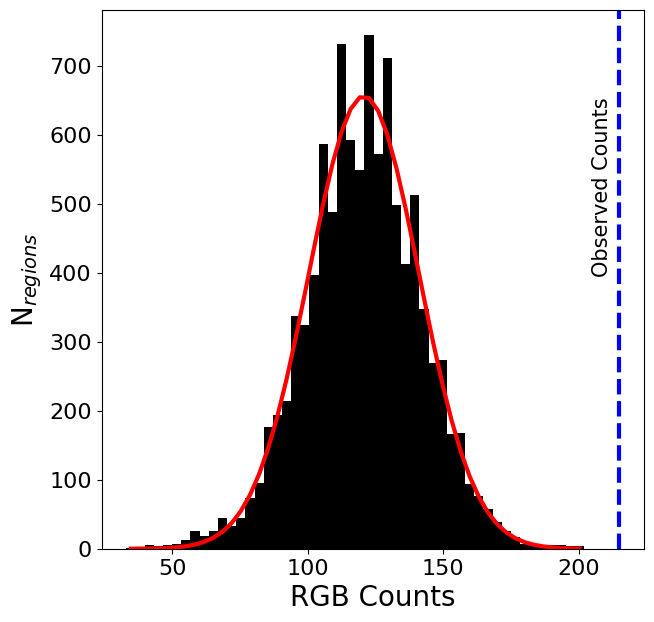}
    \caption{The distribution of RGB counts in $10^4$ empty patches of sky covering the same area as the stream selection. The best-fit normal distribution is shown in red, and the counts we detect in the stream is denoted as the blue dashed line. }
    \label{fig:bg-char}
\end{figure}

In Figure~\ref{fig:bg-char}, we show the distribution of RGB counts in background regions covering the same area as the stream selection. Using the best-fit Gaussian distribution, we determine the probability that our detection arises from a random background fluctuation to be $5.2 \times 10^{-6}$.

\begin{figure}
    \centering
    \includegraphics[width=\linewidth]{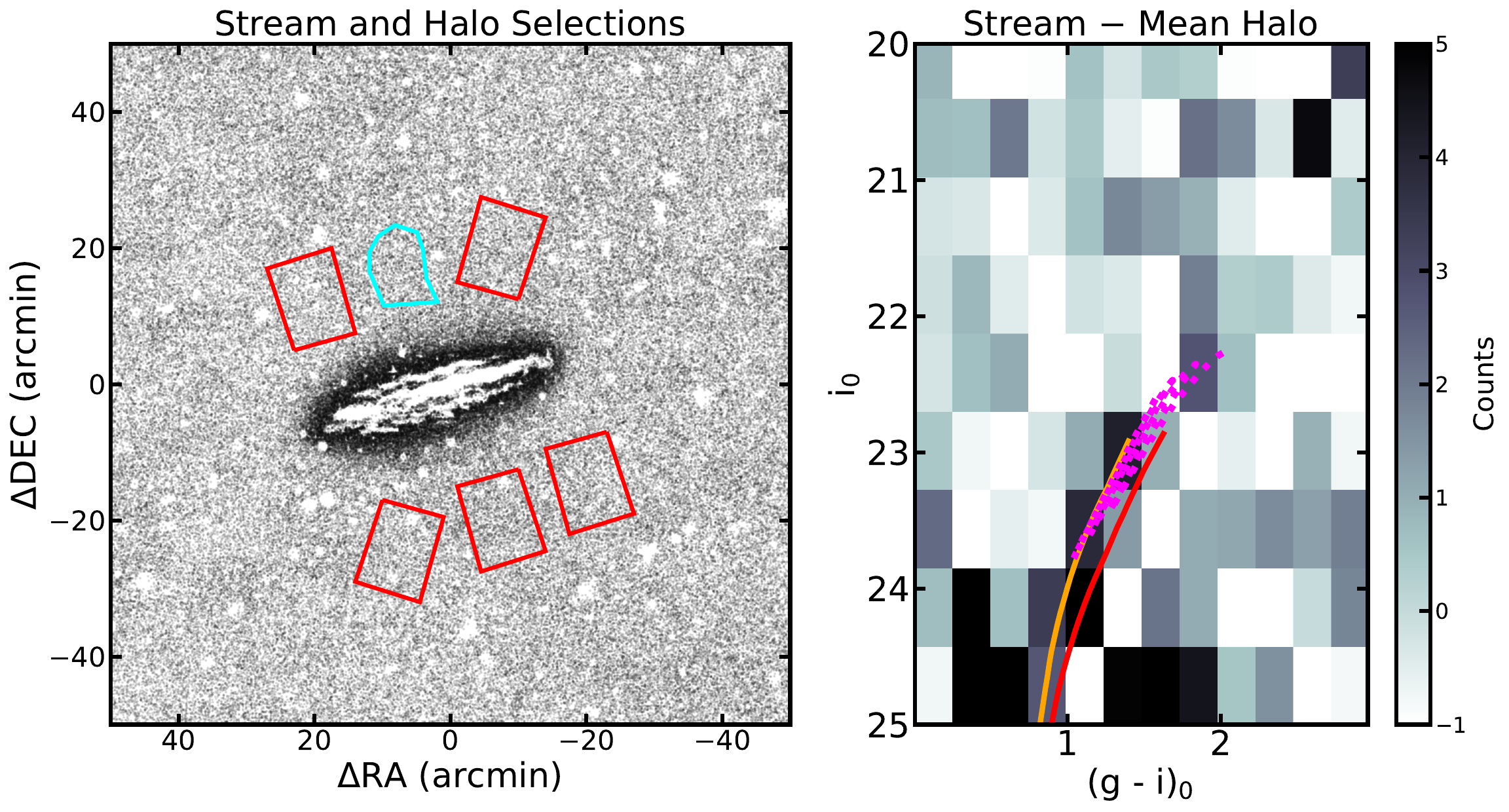}
    \caption{Left: RGB stars plotted in black, with the spatial selection of the stream (cyan) and five regions in the stellar halo of NGC~55 (red) shown. Right: The Hess diagram of the stream selection with the mean of the five halo selections subtracted. The signal we detect in Figure \ref{fig:hess} is still present. }
    \label{fig:halo-char}
\end{figure}

We also test the significance of the stream versus other regions of the stellar halo. Figure \ref{fig:halo-char} shows the selections in the halo (red) we compare with the stream (cyan) on the left, and the Hess diagram of the stream minus the mean Hess diagram of the halo selections. The RGB signal we see in \ref{fig:hess} is still visible. 

\section{Details of Archival H$\alpha$ Observations}
\label{appendix:rgb}

In Table \ref{Tab:rgb-obs} we detail the equipment and exposure times used to create the RGB image shown in Figure \ref{fig:kde-hi}. All of these observations constitute 223 hours of integration. 

\begin{deluxetable*}{ccccc}[hbt]
\tablecaption{
    \textnormal{Details of the observations used to create the three-color image in Figure \ref{fig:kde-hi}.}
    \label{Tab:rgb-obs}
}
\tablecolumns{2}
\setlength{\extrarowheight}{4pt}
\tablewidth{\linewidth}
\tabletypesize{\small}
\tablehead{
\colhead{Telescope} & 
\colhead{Observer(s)} & 
\colhead{Filters} &
\colhead{Int. Times (Hrs.)} &
\colhead{Total Int. Time (Hrs.)}
}
\startdata
\hline
24” PlaneWave CDK (Sbig 16803) & SWOS Group & LRGBH$\alpha$ & 8, 5, 5, 5, 10 & 43 \\
24” PlaneWave CDK (Moravian C5) & SWOS Group & LRGBH$\alpha$O3S2 & 13, 4, 4, 4, 28, 28, 20 & 101 \\
24” PlaneWave CDK (QHY600M) & Matt Dietrich & RH$\alpha$ O3& 4, 25, 25 & 54 \\
1m ASA RC-1000 (Chilescope T1, FLI PL 16803) & Alaxander Zaytsev & LRGBH$\alpha$O3S2 & 2, 2, 1.5, 2, 7, 5.5, 5.5, 5.5, 5.5 & 25 \\
\enddata

\tablecomments{Image Data: Hanson, Dietrich, Zaytsev, Mazlin, Parker, Forman, Magill; https://www.hansonastronomy.com/ngc-55
}
\end{deluxetable*}

% Grand Total Hours: 223 Hours

% Image Data: Hanson, Dietrich, Zaytsev, Mazlin, Parker, Forman, Magill

% \end{appendix}

% -------------
\begin{acknowledgements}

Financial support for this publication results from grant SA-LSST-2024-106c from Research Corporation for Science Advancement. BMP acknowledges support from NSF grant AST2508745. Any opinions, findings, and conclusions or recommendations expressed in this material are those of the author(s) and do not necessarily reflect the views of the National Science Foundation. ADD acknowledges support from STFC grants ST/Y002857/1. JJ acknowledges support from the National Aeronautics and Space Administration (NASA) under Grant No. 80NSSC25K0365. SP was supported by a research grant (VIL53081) from VILLUM FONDEN. SP was also co-funded by the European Union (ERC, BeyondSTREAMS, 101115754) grant. Views and opinions expressed are however those of the author(s) only and do not necessarily reflect those of the European Union or the European Research Council. Neither the European Union nor the granting authority can be held responsible for them.

% The DELVE Survey gratefully acknowledges support from Fermilab LDRD (L2019.011), the NASA Fermi Guest Investigator Program Cycle 9 (No. 91201), and the National Science Foundation (AST-2108168, AST-2307126). This work was supported in part by the U.S. Department of Energy, Office of Science, Office of Workforce Development for Teachers and Scientists (WDTS) under the Science Undergraduate Laboratory Internships Program (SULI).

The DELVE Collaboration gratefully acknowledges support from Fermilab LDRD L2019-011, the NASA Fermi Guest Investigator Program Cycle 9 No. 91201, and the National Science Foundation under Grant No. AST-2108168, AST-2108169, AST-2307126, and AST-2407526. This research is partially funded by a generous gift from Charles Simonyi to the NSF Division of Astronomical Sciences. The award is made in recognition of significant contributions to Rubin Observatory’s Legacy Survey of Space and Time

This project used data obtained with the Dark Energy Camera (DECam), which was constructed by the Dark Energy Survey (DES) collaboration. Funding for the DES Projects has been provided by the DOE and NSF (USA), MISE (Spain), STFC (UK), HEFCE (UK), NCSA (UIUC), KICP (U. Chicago), CCAPP (Ohio State), MIFPA (Texas A$\And$M), CNPQ, FAPERJ, FINEP (Brazil), MINECO (Spain), DFG (Germany), and the Collaborating Institutions in the Dark Energy Survey, which are Argonne Lab, UC Santa Cruz, University of Cambridge, CIEMAT-Madrid, University of Chicago, University College London, DES-Brazil Consortium, University of Edinburgh, ETH Zürich, Fermilab, University of Illinois, ICE (IEEC-CSIC), IFAE Barcelona, Lawrence Berkeley Lab, LMU München, and the associated Excellence Cluster Universe, University of Michigan, NOIRLab, University of Nottingham, Ohio State University, OzDES Membership Consortium, University of Pennsylvania,  University of Portsmouth, SLAC National Lab, Stanford University, University of Sussex, and Texas A$\And$M University.

\end{acknowledgements}

\software{ \texttt{NumPy} \citep{van_der_walt_numpy_2011,harris_array_2020}, \texttt{Matplotlib} \citep{hunter_matplotlib_2007}, \texttt{Astropy} \citep{robitaille_astropy_2013,collaboration_astropy_2018,collaboration_astropy_2022}}, \texttt{AGAMA} \citep{vasiliev_agama_2019}, \texttt{GADGET-4} \citep{springel_simulating_2021}, \texttt{yt} \citep{turk_yt_2011}, \texttt{WebPlotDigitizer} \citep{rohatgi_webplotdigitizer_2024}

\bibliography{refs}
\bibliographystyle{aasjournalv7.1}

\end{document}